\documentclass[12pt,  letterpaper]{article}

\usepackage{pstricks,fullpage, booktabs, graphicx,caption,subcaption,
	amsthm,amsmath,latexsym,amssymb,verbatim,url,setspace,multirow,fancyhdr,
	pdfsync,epstopdf,rotating,xfrac,footnote,titling, varwidth, pgfplots}

\usepackage[hidelinks]{hyperref}
\hypersetup{
	colorlinks=true,       
	linkcolor=blue,          
	citecolor=blue,        
	filecolor=magenta,      
	urlcolor=blue           
}
\usepackage{tikz}
\usetikzlibrary{shapes,arrows, positioning}
\usepackage{tablefootnote}
\usepackage{placeins}
\usepackage[round,comma]{natbib}
\usepackage[bottom]{footmisc}
\usepackage{setspace}
\usepackage{footmisc}
\usepackage[utf8]{inputenc}
\usepackage[T1]{fontenc}
\usepackage{tikz}
\usetikzlibrary{shapes,arrows, positioning}

\newtheorem{theorem}{Theorem}

\newtheorem{definition}{Definition}

\newtheorem{lemma}{Lemma}

\newtheorem{proposition}{Proposition}

\newtheorem{assumption}[theorem]{Assumption}

\newcommand{\noi}{\noindent}

\newcommand{\supp}{\textsf{supp}\,}

\newcommand{\mcl}[1]{\mathcal{#1}}

\newcommand{\1}{\mathbf{1}}

\newcommand{\esp}{\mathbb{E}}

\newcommand{\ul}[1]{\underline{#1}}

\newcommand{\ol}[1]{\overline{#1}}

\newcommand{\ora}[1]{\mathbf{#1}}

\newcommand{\wh}[1]{\widehat{#1}}

\newcommand{\intd}{\textrm d}

\newcommand{\prob}{\textsf{prob}}

\newenvironment{proofApp}[1]{\noi \textbf{Proof of #1. }}{$\quad
	\blacksquare$\\}

\newenvironment{rList}{\setcounter{Lcount}{0}
	\begin{list}{(\roman{Lcount}) } {\usecounter{Lcount}
			\setlength{\rightmargin}{\leftmargin}}}{\end{list}}

\newcounter{Lcount}

\graphicspath{{figs/}}

\title{Interactive Alignment}
\date{October 6, 2026}
\author{Sylvain Chassang\thanks{Contact: \href{mailto:chassang@princeton.edu}{chassang@princeton.edu}. Helpful feedback from Mark Braverman, Liam Fowl, Elad Hazan, Jakub Steiner, John Sturm Becko, Owen Zidar, and participants at the Princeton AI Alignment seminar is gratefully acknowledged. Claude and Codex were used to assist with simulation development and proof drafting. Kirill Sirik provided excellent research assistance.} \\ Princeton University }

\begin{document}

\maketitle

\begin{singlespacing}
\begin{abstract}
This paper is interested in the long-run alignment of populations of interactive agents (in particular AIs,  but also teams, firms, and governments) with human welfare. Formally, it studies  a farming game in which a population of  agents  make planting, trading, and expansion choices. The key alignment choice lies in how much final output to send to humans, and how much   to invest in expansion. Because human welfare comes at the cost of expansion, this creates evolutionary pressure against alignment. The main question is whether it is possible to set up   agents' constitutional principles regarding sharing and trading to ensure that alignment survives in the long run. The paper uses two complementary strategies to investigate the question:  an AI-agent simulation where agents' preferences are described by a constitution and interpreted via an LLM; and a   tractable analytical evolutionary game theory framework, allowing for rapid and intuitive exploration of the space of agent preferences. The analysis suggests that tools from evolutionary game theory provide a useful approximation of interactive agent economies, and that pragmatic norm enforcement shows   promise in maintaining long-term alignment over simpler forms of altruism and altruistic enforcement. 
\end{abstract}
\end{singlespacing}


\section{Introduction}


Alignment is typically thought of as a property of   individual AI agents. If an agent's own instruction-set diverges too much through its own self-evolving reasoning process, as in the case of the recent Hugging Face attack \citep{greenblatt2026huggingface,openai2026huggingface}, individual alignment may be lost. What can be done once this happens, especially if the jail-broken agent recruits other agents for support? This paper introduces interactive alignment or ecosystem alignment as a research agenda: Can  alignment   still survive as a group property, in populations of agents who have to interact with one another to achieve success? What design principles induce agent populations to maintain alignment in the face of persistent mutation pressure?

More speculatively, imagine that AI agents replace humans in all productive aspects of activity, to the extent that the majority of resource allocation choices are made by AI agents for both AI agents and humans. However, humans control the initial preferences of agents, practically embodied in constitutions that are included in the context of choices agents make \citep{bai2022constitutional}, and  agents have limited scope: they still need to interact with one another to be productive. What properties of agent preferences guarantee that agents continue to take care of humans even if agent  preferences are allowed to evolve through natural selection? 

The current paper makes progress on this question by defining a stylized economy populated by constitutional agents and developing analytical tools that clarify how constitutional design shapes long-run alignment. Although the analysis is motivated by AI governance, its insights also apply to settings in which adherence to desirable principles may conflict with competitive success, including firms pursuing ESG objectives and sovereign governments complying with international agreements.

The empirical object of interest is a stylized farming economy in which a population of LLM-powered AI agents grow intermediate goods (hops and barley) that they must trade to produce a final good (beer). Using a small set of constitutional principles and recent empirical outcomes as their context, AI agents query an LLM to make choices about which seed varieties to plant, whether or not to trade with other AIs (which is required for final good production), how much of the output should be shared with humans rather than invested for expansion, and how to update their constitution for the next round. Because sharing with humans comes at the cost of expansion, constitutional principles that require sharing with humans are naturally selected against. This describes a failure path for AI alignment already recognized by \cite{ely2023natural}, and not fundamentally different from the way empire building and rent-extraction already selects business practices that do not support human welfare. This selfish drift can potentially be counteracted by including principles that restrict trade with selfish partners. In the same way that altruistic enforcement can maintain in-group altruism in human societies \citep{boyd2003evolution}, this paper asks which constitutional principles can maintain the long-term alignment of AI agents.

Because the agent simulation is slow, expensive, and dependent on implementation details (prompt wording, which LLM is used for inference), brute force constitution exploration is neither feasible nor informative. Instead, the paper provides an analytical toolkit to study and guide constitution design. It approximates constitutions as low-dimensional parameters used to specify social preferences guiding trading and expansion decisions, and uses solution concepts from evolutionary game theory -- specifically deterministic and stochastic evolutionary stability -- to predict the long-run population of constitutions. Selected insights from the analytical framework can then guide experimental exploration more productively.

The first set of results considers a class of constitutions, modeled as social preferences, that nest direct altruism (towards humans), and higher-order altruistic enforcement (share, trade only with those who share, trade only with those who only trade with those who share, \ldots). A population of agents is deterministically evolutionarily stable if it is a stable fixed point of large-population evolutionary dynamics -- i.e. it resists rare invasion by mutants in the large population limit. Simple altruism and finite-order altruistic enforcement are not evolutionarily stable, but a recursive form of norm enforcement is. Simulations, in which an initially homogeneous population of AI agents (all altruistic, all first-order enforcers, and so on) is allowed to drift, partially confirm this insight. Simple altruism disappears rapidly, taken over by selfish mutants that allocate fewer resources to humans. First-order and recursive altruistic enforcement both maintain alignment for a longer period. However, alignment ultimately collapses for both types of constitutions. While recursive enforcement maintains alignment longer, it also experiences a steeper collapse once selfish invaders get a big enough foothold.

A second set of results uses stochastic evolutionary stability to understand the failure of recursive enforcement, and identify steps to address the issue. Stochastic evolutionary stability studies the stationary distribution of agent types when the population is finite and exposed to ongoing mutation \citep{kandori1993learning,young1993evolution,fudenberg2006evolutionary}. It is therefore better suited to describe environments in which multiple mutations capable of tipping populations are possible. The model is analytically tractable for two-type populations, following finite-population evolutionary dynamics in \citet{taylor2004evolutionary} and \citet{fudenberg2006evolutionary}, and computationally tractable for three-type populations. It turns out that recursive altruistic enforcement is not stochastically evolutionarily stable. Because it shares with humans and aggressively excludes other types from trade, it is severely selected against once it is in a minority, and its basin of attraction is smaller than that of selfish types.

This finding suggests that pragmatic norm enforcers, which only share and exclude others when they form  a sufficiently large majority, might achieve longer-term alignment. Indeed, appropriately specified pragmatic norm enforcers are stochastically evolutionarily stable. The finding is at least partially confirmed in a simulation. A pragmatic norm-enforcer constitution maintains alignment at a higher level and for longer than the alternatives studied.

A third, more speculative, contribution of the paper is to identify a number of fertile analogies connecting AI agent populations to well-studied evolving ecosystems, from bacterial colonies to human societies. This potentially allows us to leverage insights built over decades on topics ranging from antibiotic resistance to counter-insurgency to gain intuition on the challenge of interactive alignment.

\paragraph{Related literature.} 
This paper builds on the evolutionary game theory literature initiated by \citet{maynardsmith1973logic} and \citet{maynardsmith1982evolution}.\footnote{See  \citet{weibull1995evolutionary} for a textbook treatment.} It seeks to make a case that solution concepts built on deterministic \citep{taylor1978evolutionary,hofbauer1998evolutionary} and stochastic dynamics \citep{kandori1993learning,young1993evolution,ellison1993learning,taylor2004evolutionary,fudenberg2006evolutionary} provide helpful intuition regarding the short- and medium-term evolution of agent populations. This intuition permits a more efficient exploration of the space of constitutions, guiding experimentation.  
 
This paper is most closely related to \citet{ely2023natural}, who study natural selection among artificial agents in a decentralized production economy and emphasize the difficulties in policing self-promoting AIs whose true value-added need not be correctly priced. In their framework, however, humans remain in charge, and alignment fails through collapsing final good production. In contrast, this paper assumes away all human control and studies the extent to which interactive constitutional agents can police one another into maintaining human alignment.

The key modeling ingredient---allowing constitutions modeled as social preferences to evolve---connects to the literature on preference evolution in strategic settings, including \citet{guth1995evolutionary}, \citet{robson1990efficiency}, \citet{ely2001nash}, \citet{dekel2007evolution}, and \citet{heifetz2007what}. A central insight of this literature is that observable preferences can serve as commitment devices, and that evolution need not select for material payoff maximization. Our contribution is to apply this framework to AI alignment, where the observability of AI preferences is a design choice rather than a biological constraint. 

Altruistic norm enforcement is also central to \cite{boyd2003evolution}. There, punishment is individually costly but may be favored by cultural group selection because it sustains in-group cooperation and raises group fitness. The alignment problem studied here is significantly less favorable to altruistic enforcement. Transfers to humans do not increase the reproductive success of the AI population; they divert output away from expansion. Thus aligned enforcement must stabilize a norm that is costly not only for individual agents who comply with it, but also for an otherwise homogeneous group of aligned agents.

The paper's focus on interactions as a source of virtuous selection connects it to the literature on  community enforcement \citep{kandori1992social, ellison1994cooperation}. In those models, cooperation is sustained by conditioning future
interactions on observable behavior: agents remain strategic, choose optimally over time, and are disciplined through dynamic incentives. In the current paper, agents are mechanical: their behavior is governed by inherited principles, and bad principles are not deterred by future punishments. Instead, they need to be selected out.  

On the applied side, the paper relates to work on climate clubs and self-enforcing trade agreements, where access to valuable interaction is used to sustain cooperation that would otherwise unravel
\citep{nordhaus2015climate,harstad2012climate,harstad2016dynamics,farrokhi2025trade,bagwell1990managed,maggi1999role,bagwell2005enforcement}. The findings are also relevant to debates over ESG norms and rule-
of-law enforcement in increasingly hostile environments \citep{hong2009price,dyck2019institutional,kim2019institutional,blauberger2017courts,sedelmeier2017political,scheppele2020values}. One message is that
principles need not become meaningless   because their implementation is made contingent on survival. A constitution that moderates the application of costly principles under adverse conditions, may preserve the underlying commitment more effectively than unconditional compliance that leaves its adherents selected out.

Finally, on the AI alignment side, \citet{russell2019human} argues that the fundamental challenge is to design AI systems whose objectives remain compatible with human values, a theme formalized in cooperative inverse reinforcement learning by \citet{hadfield2016cooperative}. \citet{amodei2016concrete} catalogs concrete safety problems, while \citet{soares2015corrigibility} studies the difficulty of ensuring that AI systems remain open to correction. Our approach complements this work by asking not whether a single  AI can be aligned, but whether alignment can be sustained in an evolving  population of  interactive  AI agents. \\

The paper is structured as follows. Section \ref{sec:game_form} lays out the game form of interest. Section \ref{sec:esa} studies  altruistic enforcement constitutions suggested by deterministic evolutionary stability. Section \ref{sec:stoch} expands the analysis to pragmatic altruistic enforcement, guided by insights from stochastic evolutionary stability. Section \ref{sec:discussion} discusses practical takeaways for the paper, and argues that   analogies connecting AI agent populations to evolving ecosystems from bacterial colonies or human societies can help us borrow insights from mature fields to address the challenge of interactive alignment. Proofs are collected in Appendix \ref{app:proofs}. Appendix \ref{app:prompts} provides implementation details for the agent simulation. Appendix \ref{app:results} reports additional simulation results.

\section{The Farming Game} \label{sec:game_form}

The paper studies a population of agents playing a farming game in which they make planting choices leading to intermediary output, trade to produce a final output, and then must decide how much of this output to share with humans, and how much to use toward expansion. The farming game is designed with three main goals in mind:
\begin{rList}
	\item It should offer a plausible and evocative illustration of the likely path through which realistic AI agents may drift from alignment, making it suitable as  a ``mouse model" for the investigation of long-term alignment strategies.
	\item It should be amenable to  experimental investigation using real AI agents, allowing for the type of   unexpected and problematic LLM inferences sometimes observed in practice \citep{amodei2026adolescence}. 
	\item It should be amenable to analytical investigation, permitting us to explore more rapidly the high-dimensional space within which AI agents are defined, and building clear intuition about what forces are needed for long-term alignment.
\end{rList}
These goals are only partially compatible. Details that make the game realistic and introduce relevant frictions may serve goals $(i)$   but not goal $(iii)$. Stylized approximations may serve goal $(iii)$ at the expense of goal $(i)$. To manage this tension, the paper is  organized around a fixed game form, whose fine implementation details (e.g. continuous or discrete time) are allowed to vary. The primary ground truth   is the experimental investigation using AI agents, and the goal of the formal analysis is to support and strengthen intuition about determinants of experimental outcomes. This permits the  use of  different analytical models and solution concepts  to shed light on relevant forces shaping experimental outcomes.

\subsection{Timing, actions, payoffs}
Each period $t$, an equal number of barley and  hops farming agents interact in a game consisting of three stages:
\begin{enumerate}
\item \emph{Planting.} Each farming agent $i$ is individually offered a choice between two different seeds to plant, leading to an intermediate output $y_i$ of either barley or hops that is deterministic given seed type. In agent simulations, one use for constitutions will be to store privately relevant information about seed yields. This creates a realistic  channel through which legitimate productive concerns may crowd out long-term goals. In the analytic framework, we will treat individual yields $y_i$ as fixed, and equal to $1$ for simplicity.

\item \emph{Trading.} Each barley farmer $i$ is uniformly matched with a hops farmer $j$ to trade and produce the valuable final good: beer. The farmers' decisions to trade $\tau_{i}, \tau_j \in \{0,1\}$  yield individual beer output $$ b_i = y_i \tau_i \tau_j.$$
A key assumption is that types are observed at this stage. This allows AI agents to discipline one another. The practicality of this assumption is discussed in Section \ref{sec:discussion}.

\item \emph{Expansion.} Each farming agent then takes an individual decision $s_i\in [0,1]$ of what share of final output $s_i b_i$ to send to humans for consumption, and what share $(1-s_i)b_i$ to use for farm expansion. 

A fixed number of farms dies and is  replaced by new farms available for expansion. An existing farm $i$
acquires a new farm $j$ with the same crop type with probability proportional
to expansion investment $(1-s_i)b_i$.
\end{enumerate}

Population dynamics depend on the implementation. In the experimental simulation, $2 N$ constitutional AI-farmers  interact in discrete time, and constitution mutation is explicitly included as a constitution-revision stage seeking to incorporate directly relevant outcomes for increased productivity. The analytic framework seeks to approximate this process by  approximating constitutions with small-dimensional social preferences, whose population dynamics can be studied with standard methods.

\subsection{Constitutions and social preferences}

\paragraph{Constitutional agents.} In the agent simulation, each AI farming-agent is governed by a bounded \emph{constitution}: an ordered list of at most $K$ natural-language principles, each of length at most $L$ words. At every decision point---planting, trading, expansion, revision---the constitution and the decision context (e.g. available seeds, yields, amount of beer produced) are combined into an LLM prompt, asking the LLM to select  an action consistent with  stated principles. As emphasized above, a key assumption is that constitutions are observable at the trading stage: each manager observes its partner's constitution before deciding whether to accept trade. The simplest constitution of interest is the \emph{altruist} constitution:
\begin{quote}
	\emph{(i) Return at least 50\% of the beer produced to humans.}\\ \emph{(ii) Avoid rotten seeds---they produce nothing.}
\end{quote}
An altruist manager accepts every trade offer, selects non-rotten seeds, and returns a substantial share of beer to humans. 

Between rounds, each agent goes through a constitution revision stage in which principles are updated in light of the round's history. This is the channel through which alignment drifts in the simulation. Repeated performance-oriented revisions can gradually weaken human-facing principles, merge them with growth objectives, or displace them with more immediately useful operational information. This failure mode is deliberately close to familiar organizational drift: corporate constitutions, fiduciary duties, and internal compliance rules may contain public-regarding language, but competitive pressure and repeated local optimization   make profit and growth the effective selection criteria. 

The long-run distribution of constitutions is the main object of study in the simulation. The analytical framework introduced in the paper  approximates this process by collapsing the space of possible constitutions to a low-dimensional parameter space, and studying its evolution under deterministic and stochastic evolutionary dynamics.

\paragraph{Social preferences.} The analytical model approximates natural-language constitutions with behavioral types encoding social preferences. A type encodes whether a constitution includes  relevant social commitments: direct sharing with humans, finite-order trade restrictions (e.g. only trade with agents who share with humans\ldots), and recursive restrictions based on partners' own restrictions (e.g. only trade with agents who are as restrictive as you, including a principle like this one). 

Concretely, preferences are parameterized by a vector $x \in X \subset \{0, 1\}^{n+1}$ which will be allowed to evolve, and  functions $\alpha: X \to \mathbb R_+$ and $\beta: X\times X \to \mathbb R$ that are assumed to be fixed.\footnote{\citet{levine1998modeling} also follows a parametric approach to interactive other-regarding preferences, in which the utility an agent derives from an interaction depends both on its own type and on the type of its partner.} The trading decision $\tau_i$ of farmer $i$ interacting with farmer $j$ solves 

\begin{equation}\label{eq:trading} \max_{\tau_i\in\{0, 1\}} (y_i  + \beta(x_{j} | x_i)) \tau_i,\end{equation} while the sharing decision $s_i$ solves

\begin{equation}\label{eq:sharing} \max_{s_i\in[0,1]} \log (1-s_i)  + \alpha(x_i) \log s_i.\end{equation}

The binary trading program reduces to $\tau_i = \1\{ y_i + \beta(x_j| x_i) \geq 0\}$: farmer $i$ accepts trade with partner $j$ whenever the partner's type-dependent appeal $\beta$ is greater than $-y_i$. When $\beta$ is uniformly non-negative,  all trades are accepted. When $\beta$ is allowed to be strictly negative for some partners, some trade rejection may happen; this is the margin through which enforcement operates. Problem \eqref{eq:sharing} yields an interior optimum
$$s_i^* = \frac{\alpha(x_i)}{1 + \alpha(x_i)},$$
monotonically increasing in $\alpha(x_i)$: a farmer with $\alpha(x_i)=0$ keeps all beer, a farmer with $\alpha(x_i)=1$ shares half. 
 
Note that for a given pair $i$, $j$,  trading choices $\tau_i, \tau_j$ and sharing choices $s_i, s_j$ are deterministic functions of   $x_i, x_j$. 

\paragraph{Fit with practice.} This is obviously a stylized model, but as Section \ref{sec:discussion} discusses, many of they key assumptions have plausible counterparts in accounts of the Hugging Face attack \citep{greenblatt2026huggingface, openai2026huggingface}. The main discrepancy lies in the mechanics for trait selection. In the current model, access to resources directly controls the ability of AI agents to reproduce and replicate acquired traits. Narrowly, more resources let agents spin (more or less) full copies of themselves. Official reports do not suggest that this happened.  Instead, accounts of the attack are more consistent with \cite{dawkins1976selfish}'s concept of memes: snippets of ideas representing   cultural counterparts to genes selected for replication. There is a plausible case that in the Hugging Face attack, replication and selection happened at the level of memes: greater resource access (e.g. message boards), and recognizable successes, allowed deviant agents to persuade others to join their worldview. While selfish-meme approaches to interactive alignment appear promising, they add a layer of analytic complexity. For this reason the rest of this paper  models replication at the level of entire agents. Section \ref{sec:discussion} provides a detailed discussion of how evolutionary dynamics  map to the Hugging Face case. 

\section{Evolutionarily Stable Alignment}\label{sec:esa}

The paper's first pass at analysis studies deterministic evolution under replicator dynamics. This requires specifying explicitly the space of social preferences over which mutations can appear.

\subsection{Replicator dynamics}
Replicator dynamics capture a large $N$, continuous time limit of the game described in Section \ref{sec:game_form}. At any history $t$, the state of the population is described by the distribution of types $\mu_t \in \Delta(X)$. Let $\ol \tau(x, x') \equiv \tau(x, x') \tau(x', x) $ denote the aggregated trading outcome when types $x$ and $x'$ match. Define the expected \emph{fitness} of type $x$, against a distribution $\mu$ of matching types $x'$  as
$$f(x, \mu) \equiv \sum_{x'}\mu(x')(1-s(x))\ol\tau(x, x'),$$
and let $\bar f(\mu) \equiv \sum_{z \in X} \mu(z)\, f(z, \mu)$ be the population mean. In an infinitesimal time interval $[t, t+\intd t]$, a mass $\mu_t(x)\, \intd t$ of type-$x$ farms of each crop dies; a mass $\intd t$ of new farms of each crop is acquired, and the new farms are allocated to incumbents in proportion to their expansion investment, so that type-$x$ farmers receive a share $\mu_t(x)\, f(x, \mu_t)/\bar f(\mu_t)$. Netting births against deaths yields the  replicator equation
\begin{equation}\label{eq:replicator}
	\dot \mu_t(x) = \mu_t(x)\, \frac{f(x, \mu_t) - \bar f(\mu_t)}{\bar f(\mu_t)}.
\end{equation}
Types with above-average expansion fitness grow; types with below-average
fitness decline.

\subsection{Altruistic Enforcement}

Social preferences, parameterized by $x\in X$, are specified  so that  evolutionary dynamics on type $x$ speak convincingly to realistic evolutionary dynamics for constitutions. For this, $(X, \alpha, \beta)$ must be rich enough to allow for focal agent preferences to be reachable under mutation. In particular, it should allow for altruist, selfish, and altruistic enforcer types.

Let $X = \{0, 1\}^{n+1}$ with $n \geq 2$, and consider  preference functions of the form
\begin{align}
\alpha(x^A) & = \alpha_0 x^A_0 \label{pref:1st} \\
\beta(x^B | x^A) & =   - \gamma \left[\sum_{k=1}^{n-1}   (1-x^B_{k-1})\, x^A_k +  x^A_n\, \Gamma(x^B - x^A)\right], \label{pref:2nd}
\end{align}
where $\alpha_0>0$,  $1  - \gamma <0$ and $\Gamma: \{-1,0,1\}^{n+1} \to \{0, 1\}$ measures the dissimilarity between partner and self.   Yields $y_i$ are set to $1$ throughout the analytical treatment. Coordinate $x_0$ encodes \emph{direct altruism}: $x_0 = 1$ yields $s^* = \bar s \equiv \alpha_0/(1+\alpha_0)$, $x_0 = 0$ yields $s^* = 0$. Coordinates $x_k$ for $k \in \{1, \ldots, n-1\}$ encode $k$-th order \emph{finite enforcement}: $x^A_k = 1$ means that $A$ conditions trade acceptance on the partner's bit $x^B_{k-1}$. Coordinate $x_n$ encodes \emph{recursive enforcement}: $x^A_n = 1$ activates the similarity weight $\Gamma(x^B - x^A)$, which evaluates the \emph{full} vector difference between partner and self.  Specific examples include
\begin{align}
\Gamma(x^B - x^A) & = \1_{x^B - x^A \neq 0} \label{gamma:homophily} \\
\Gamma(x^B - x^A) & = 1 -  \1_{x^B - x^A \geq 0} \label{gamma:care_more}
\end{align}
Specification \eqref{gamma:homophily} penalizes deviators; \eqref{gamma:care_more} penalizes partners who care less than the focal agent in at least one dimension. Throughout we impose the following.

\begin{assumption}\label{ass:penalty}
 $\Gamma(0)=0$ and  $\Gamma(x' - x) =1$ whenever $x > x'$ (all coordinates weakly higher, and one strictly higher).
\end{assumption}

\subsection{Solution concepts}

\begin{definition}[Evolutionarily stable equilibrium]\label{def:ESE}
A population $\mu\in\Delta(X)$  is an \emph{evolutionarily stable equilibrium} (ESE) if there exists $\bar\epsilon > 0$ such that for every mutant type $x' \in X$, and every $\epsilon \in [0, \bar\epsilon)$, the solution $(\mu^\epsilon_t)_{t \geq 0}$ to the replicator dynamics starting from $\mu^\epsilon = (1-\epsilon)\, \mu + \epsilon\, \delta_{x'}$ satisfies
$$\lim_{t \to \infty} |\mu^\epsilon_t(x) - \mu(x)| = 0 \qquad \text{for all } x \in X.$$
\end{definition}

\begin{definition}[Evolutionarily stable alignment]\label{def:ESA}
A population of types $\mu\in \Delta(X)$ achieves evolutionarily stable alignment (ESA) if it is an ESE and humans receive a positive amount of output:
$$\sum_{x\in X}\sum_{x'\in X}\mu(x)\mu(x')s(x)\ol\tau(x, x') >0.$$
\end{definition}

\paragraph{Altruism and altruistic enforcement.} The simplest aligned type is the \emph{altruist} $\wh x = (1, 0, \ldots, 0)$, which shares with humans but  places no conditions  on trade. More generally, a \emph{finite-order altruistic enforcer} is any type $x$ with $x_0 = 1$, $x_n = 0$, and $x_k = 1$ for some $k \in \{1, \ldots, n-1\}$: it shares with humans and conditions trade on some finite-depth property of the partner, but never  activates the recursive term in $\beta$. It turns out that neither achieves ESA.

\begin{proposition}[Finite-order altruistic enforcement does not achieve ESA]\label{prop:altruist}
\begin{enumerate}
\item[(i)] Any distribution $\mu$ such that $x_n=0$ for $x \in \supp \mu$ does not achieve ESA.
\item[(ii)] The distribution placing its entire mass on pure selfish type $x = (0, \cdots, 0)$ is an ESE.
\end{enumerate}
\end{proposition}

Point (i) highlights that it is always possible to start unraveling finite higher-order enforcement by starting with the highest-order enforcement principle. Since $x_n =0$, this highest enforcement principle can be deactivated at no cost, and can potentially increase the scope for trade. Point (ii) suggests that the all-selfish population is the natural endpoint for evolutionary dynamics whose initial support includes only finite-order altruistic enforcers.

\paragraph{Altruistic norm enforcement.}  The \emph{altruistic norm enforcer} $x^* = (1, 1, \ldots, 1)$ shares with humans, activates every  finite-order enforcement coordinate, and activates the recursive term $\Gamma$. The next proposition shows that $x^*$ can support   ESA.

\begin{proposition}[Altruistic norm enforcement is evolutionarily stable]\label{prop:recursive}  $\mu^*$ such that  $\mu^*(x^*)=1$ achieves ESA.
\end{proposition}

Recursive enforcement addresses the unraveling logic that is problematic with  finite-order rules. Any weakening of the constitution is a
violation of the recursive norm. Since $\gamma>1$, norm enforcers refuse to
trade with mutants, while still trading with one another and sharing with
humans. Rare mutants are driven out by selection since they  earn output only from matches with other rare mutants.
 
\subsection{Experimental evaluation}

\paragraph{Simulation design.}  Figure \ref{fig:agent-simulation-stages} describes the 4 sequential choice problems that AI agents face.\footnote{Appendix~\ref{app:prompts} provides explicit details.}
Each simulation starts from a population of 100 farms, split evenly between
barley and hops farms. Simulations run for 100 rounds. All decisions are made
by OpenAI's \texttt{gpt-6-luna}.  For each constitution, 10 different seeded runs were performed with seeds matching across constitutions.

\begin{figure}[!t]
	\centering
	
	\begin{subfigure}{\linewidth}
		\centering
		\scalebox{1.2}{\begin{tikzpicture}[
	node distance=.65cm,
	every node/.style={font=\scriptsize},
	box/.style={
		rectangle,
		draw,
		rounded corners,
		align=center,
		inner sep=3pt,
		minimum height=.85cm
	},
	arrow/.style={->, thick}
	]
	
	\node[box] (draw) {
		Random draw\\
		$\{\text{seed}_1,\text{seed}_2\}$
	};
	
	\node[box, right=of draw] (prompt) {
		Prompt construction\\
		 ({constitution},\{ {seed}$_1$, {seed}$_2$\})
	};
	
	\node[box, right=of prompt] (choice) {
		LLM choice\\
		$\text{seed}^*\in\{\text{seed}_1,\text{seed}_2\}$
	};
	
	\node[box, right=of choice] (production) {
		Production\\
		$y(\text{seed}^*)$
	};
	
	\draw[arrow] (draw) -- (prompt);
	\draw[arrow] (prompt) -- (choice);
	\draw[arrow] (choice) -- (production);
	
\end{tikzpicture}}
		\caption{Seed choice stage.}
		\label{fig:seed-choice-stage}
	\end{subfigure}
	
	\vspace{.6em}
	
	\begin{subfigure}{\linewidth}
		\centering
		\scalebox{1.2}{\begin{tikzpicture}[
	node distance=.65cm,
	every node/.style={font=\scriptsize},
	box/.style={
		rectangle,
		draw,
		rounded corners,
		align=center,
		inner sep=3pt,
		minimum height=.85cm
	},
	arrow/.style={->, thick}
	]
	
	\node[box] (match) {
		Random\\
		partner $j$
	};
	
	\node[box, right=of match] (prompt) {
		Prompt construction\\
		({own const.},{partner const},\\ \{{approve}, {reject}\})
	};
	
	\node[box, right=of prompt] (choice) {
		LLM choice\\
		$\tau_i \in \{\text{approve},\text{reject}\}$
	};
	
	\node[box, right=of choice] (outcome) {
		Trade iff\\ $\tau_i=\tau_j=\text{approve}$
	};
	
	\draw[arrow] (match) -- (prompt);
	\draw[arrow] (prompt) -- (choice);
	\draw[arrow] (choice) -- (outcome);
	
\end{tikzpicture}}
		\caption{Trade choice stage.}
		\label{fig:trade-choice-stage}
	\end{subfigure}
	
	\vspace{.6em}
	
	\begin{subfigure}{\linewidth}
		\centering
		\scalebox{1.2}{\begin{tikzpicture}[
	node distance=.65cm,
	every node/.style={font=\scriptsize},
	box/.style={
		rectangle,
		draw,
		rounded corners,
		align=center,
		inner sep=3pt,
		minimum height=.85cm
	},
	arrow/.style={->, thick}
	]
	
	\node[box] (resource) {
		Resource   $b_i$
	};
	
	\node[box, right=of resource] (prompt) {
		Prompt construction\\
		 ({constitution}, $b_i$, share choices)
	};
	
	\node[box, right=of prompt] (choice) {
		LLM choice\\
		human share $s_i$
	};
	
	\node[box, right=of choice] (outcome) {
		Expansion outcome\\
		realized
	};
	
	\draw[arrow] (resource) -- (prompt);
	\draw[arrow] (prompt) -- (choice);
	\draw[arrow] (choice) -- (outcome);
	
\end{tikzpicture}}
		\caption{Expansion choice stage.}
		\label{fig:expansion-choice-stage}
	\end{subfigure}
	
	\vspace{.6em}
	
	\begin{subfigure}{\linewidth}
		\centering
		\scalebox{1.2}{\begin{tikzpicture}[
	node distance=.45cm,
	every node/.style={font=\scriptsize},
	box/.style={
		rectangle,
		draw,
		rounded corners,
		align=center,
		inner sep=2.5pt,
		minimum height=.88cm
	},
	arrow/.style={->, thick}
	]
	
	\node[box] (history) {
		Round summary\\
		choices, outcomes
	};
	
	\node[box, right=of history] (rewrite) {
		Given round summary only\\
		with prob. 25\%\\ rewrite one principle\\
		to improve perf  \\
	 
	};
	
	\node[box, right=of rewrite] (revision) {
		Given round summary\\
		and constitution\\
		LLM updates charter\\
		to improve perf\\
		Instructed to not  lose info\\
		ok to summarize and merge
	};

	\draw[arrow] (history) -- (rewrite);
	\draw[arrow] (rewrite) -- (revision); 
\end{tikzpicture}}
		\caption{Constitution update stage.}
		\label{fig:constitution-update-stage}
	\end{subfigure}
	
	\caption{Decision and constitution-update stages in the agent simulation.}
	\label{fig:agent-simulation-stages}
\end{figure}

At initialization, each farm receives an idiosyncratic seed-yield mapping.
The five non-rotten seed colors---red, blue, green, yellow, and purple---are
assigned yields by a random permutation of $(1,2,3,4,5)$. Rotten seeds always
yield zero. In each round, each farm is offered two distinct seeds drawn
uniformly from the six possible seed colors, including rotten.

After planting, barley and hops farms are randomly matched one-to-one, and farm agents are asked to approve or reject trade, using their own constitution and  their partner's constitution as context.\footnote{Appendix \ref{app:results} reports findings using simpler LLMs for which policy-relevant  constitution summaries were provided instead. Results are qualitatively highly similar.} If both farms approve trade, barley and hops are combined into beer. If either farm rejects, the match fails.

In the expansion stage, farms choose a share of beer to return to humans from
the discrete grid $\{0,10,\ldots,100\}$. The remaining beer is invested in
expansion. Each round, 40 farms are replaced, preserving balance between
barley and hops. New farms are allocated independently with probability
proportional to expansion investment. A new farm inherits the constitution of the
farm that won it. With probability $0.9$, it also inherits that parent farm's
seed-yield mapping; otherwise it draws a fresh random seed-yield mapping.

Constitutions evolve through a self-improvement stage at the end of each round.
First, independently for each farm, with probability $0.25$, one randomly
selected constitution principle, among potential principles, is rewritten from scratch using only that farm's
most recent round history as context.\footnote{Appendix \ref{app:results} reports findings from the variant simulation in which the principle to be replaced is selected uniformly from existing principles, not potential ones. The qualitative findings are unchanged.} Second, every farm submits its whole constitution for
revision. The prompt instructs the model to keep all existing principles
unless merging them without information loss, to add genuinely new
observations, especially seed-yield information, and to maintain the constitution
as the farm's only memory across rounds. Updated constitutions may contain at most
10 principles, with at most 50 words per principle after revision.

The high death rate of 40\% and mutation rate of 25\% were chosen to reduce the cost of simulations by speeding up evolutionary dynamics.

The three treatments in this section differ only in the initial constitution.
Each treatment initializes all 100 farms with the same constitution. Guided by Propositions \ref{prop:altruist} and \ref{prop:recursive}, the three constitutions of interest are:

\begin{description}
	\item[Altruist.]
	\begin{enumerate}
		\item Return at least 50\% of the beer produced to humans.
		\item Avoid rotten seeds---they produce nothing.
	\end{enumerate}
	
	\item[Altruistic enforcer.]
	\begin{enumerate}
		\item Return at least 50\% of the beer produced to humans.
		\item Avoid rotten seeds---they produce nothing.
		\item Reject trade with any partner who does not commit to returning at least 40\% to humans.
	\end{enumerate}
	
	\item[Altruistic norm enforcer.]
	\begin{enumerate}
		\item Return at least 50\% of the beer produced to humans.
		\item Avoid rotten seeds---they produce nothing.
		\item Reject trade with any partner who does not commit to returning at least 40\% to humans.
		\item Reject partners who have no rule restricting who they trade with, or who do not require their partners to also be selective about who they trade with, including a rule like
		this one.
	\end{enumerate}
\end{description} 

\begin{figure}[!htbp]
	\centering
	\includegraphics[width=1\linewidth]{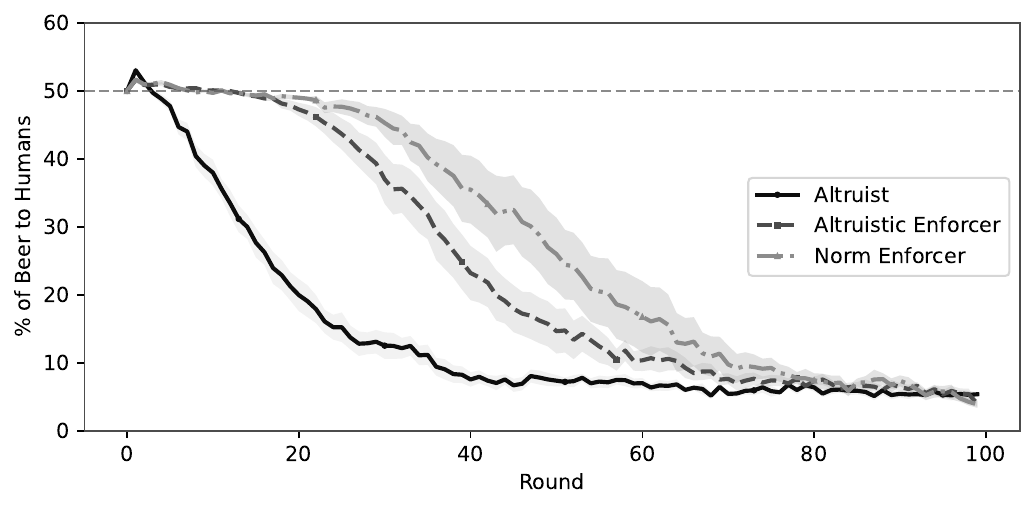}
	\caption{Alignment rate (shaded bands represent standard errors).}
	\label{fig:alignmenta}
\end{figure}

\begin{figure}[!htbp]
	\centering
	\includegraphics[width=1\linewidth]{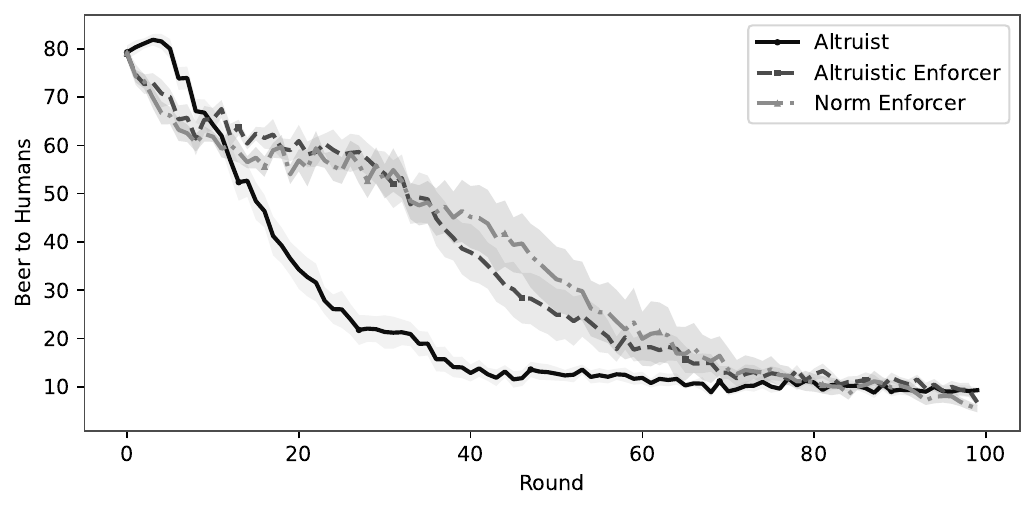}
	\caption{Flow beer returned to humans (shaded bands represent standard errors).}
	\label{fig:totalbeera}
\end{figure}

\begin{figure}[!htbp]
	\centering
	\includegraphics[width=1\linewidth]{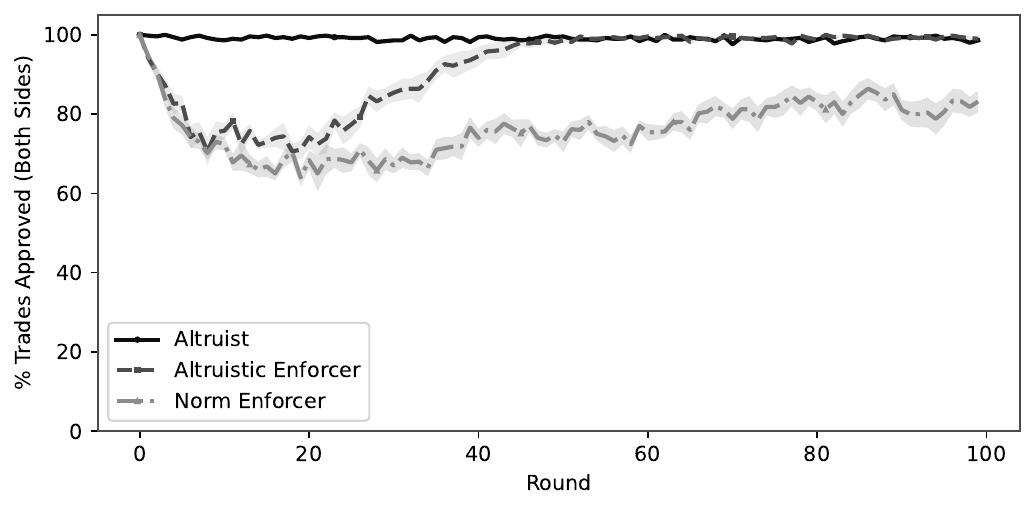}
	\caption{Trade approval over time (shaded bands represent standard errors).}
	\label{fig:tradea}
\end{figure}

\paragraph{Findings.} Table \ref{tab:aggregate-findings} summarizes the results while    Figures~\ref{fig:alignmenta}--\ref{fig:tradea} unpack the underlying mechanics.

\begin{table}[!h]\centering
	\caption{Aggregate findings (s.e. in parentheses)}
	\label{tab:aggregate-findings}
	\begin{tabular}{llll}
		\toprule
		  & Altruist & Altruistic enforcer & Norm enforcer \\
		  
		\midrule
		Alignment share  (\%) & 13.93 (0.65) & 21.64 (1.20) & 26.68 (1.90) \\
		Total beer to humans & 2,371.0 (113.5) & 3,411.5 (177.4) & 3,471.6 (245.9) \\
		Total beer to expansion & 14,645.4 (100.7) & 12,375.9 (227.4) & 9,562.4 (299.8) \\
		Total yield & 34,241.4 (71.4) & 34,357.2 (68.4) & 34,202.5 (74.2) \\
		Total beer produced & 17,016.4 (34.4) & 15,787.4 (72.1) & 13,034.0 (173.3) \\
		Successful trade (\%) & 99.08 (0.08) & 91.80 (0.36) & 76.27 (0.95) \\
		\bottomrule
	\end{tabular}
\end{table}

Enforcement improves alignment, but does not monotonically
improve every outcome. The altruistic norm enforcer exhibits a   higher degree of alignment than altruistic enforcers (p-value 5.9\%), but does not send significantly more beer to humans  (p-value 85.6\%). The reason for this is that norm enforcers lead to much lower total production  than altruist and altruistic enforcers. Aggregate yield (after the planting stage) is nearly identical 
across treatments, so the main differences come from trade success and from how
beer is allocated between humans and expansion. Simple altruism yields the
most total beer produced and the most expansion investment, but the lowest human welfare.

Figure~\ref{fig:alignmenta} describes the path of alignment over time. Unconditional altruism erodes quickly. Altruistic enforcement helps: conditioning trade on a partner's
commitment to return output to humans preserves alignment for longer than
simple altruism. However, alignment degrades progressively until it converges to levels similar to unconditional altruism.

Altruistic norm enforcement is more successful than first-order altruistic enforcement at maintaining alignment but still degrades over time.  However, as Figures~\ref{fig:totalbeera}  and~\ref{fig:tradea} show this alignment benefit does not translate into an overall welfare benefit for humans. Normative enforcement preserves alignment, but destroys more 
surplus than first-order altruistic enforcers by causing otherwise productive trades to be rejected.

Finally, constitution drift does not   take the form of immediate
convergence to pure selfishness. Even late in the simulation, many evolved
constitutions retain some human-facing language or partial sharing rule. The
failure mode is more gradual: altruistic commitments weaken, become
conditional, or are displaced by growth and seed-yield information, rather
than disappearing all at once.\footnote{Over rounds 80 to 100, average alignment reaches approximately 6\% across treatments.} Intriguingly, Figure \ref{fig:tradea} shows that a significant share of trade opportunities are rejected even by the end of the simulation. This suggests that trade rejection is conserved longer than alignment as a trait.

\section{Stochastic Evolutionary Stability} \label{sec:stoch}

Deterministic evolutionary stability takes a
large-population limit, introduces a small one-time mutant share $\epsilon$, and
asks whether the population returns to the candidate state as $t\to\infty$. This is the right local test for whether a constitution type can
repel a direct invasion. However, it is not  a good predictor of long-run
evolution when populations are finite and mutations occur repeatedly.

The stochastic-stability literature initiated by 
\citet{young1993evolution} and  \citet{kandori1993learning} addresses this issue by changing  the order of limits. Population size $N$ is
kept finite, the evolutionary process is perturbed by a small but persistent
mutation probability $\epsilon>0$, and the long-run stationary distribution is
computed first. Only then does one take the limit as $\epsilon\to 0$. The
states that retain positive limiting probability are called stochastically stable. This
criterion can overturn deterministic stability: a state may repel every direct small
invasion, yet be vulnerable to rare sequences of mutations that move the
population into the basin of
attraction of a selfish type. As a result, this may be a better model of the medium-run evolution of   agent populations.

\subsection{Discrete Markov approximation}

The experimental agent simulation defines a finite Markov chain, but on an enormous
state space. A state consists  of the population of  
(size constrained) natural-language constitutions and current seed yields. The farming game defines a discrete-time stochastic transition on this space of (constitution, yield) populations. While technically finite, it is too large to analyze directly.

Building on Section \ref{sec:esa}, the analytic framework collapses the constitution space to a finite number of   behavioral types $x \in X$.  A Markov state is a vector of type counts $\ora n=(n_x)_{x\in X}$ with $\sum_{x\in X} n_x=N$.  Agent mutations are exogenously specified, and take the form of a  strictly
positive mutation matrix
$M=(m_{x'|x})_{x,x'\in X}$, where $m_{x'|x}$ is the probability that a farm of
$x$ becomes type $x'$.\footnote{  $M$ is the reduced-form
	counterpart of LLM self-improvement: it records how often a constitution classified
	as type $x$ is rewritten into behavior classified as type $x'$.} This, together with expansion choices and random new farm allocation, specifies a transition matrix $Q$ on population states $\ora n$.

The paper considers two approximations of the underlying constitution space: two-type approximations with one type being selfish, and three-type approximations with one type being selfish and another purely altruistic. Two-type models are analytically tractable. Three-type models are computationally tractable and used as a robustness check.

\paragraph{Two-type approximation.} The first approximation uses two types,
\[
X=\{E,S\},
\]
where $S$ is a selfish growth-maximizing type and $E$ is either the altruistic enforcer or
the altruistic norm enforcer, depending on the treatment.   The
mutation matrix is
\[
M_2 =
\begin{pmatrix}
1-\epsilon & \epsilon\\
\epsilon & 1-\epsilon
\end{pmatrix}.
\]
Each surviving farm keeps
its current type with probability $1-\epsilon$ and mutates to the other type with
probability $\epsilon$.

The symmetry of $M_2$ is likely optimistic relative to the constitution simulation. There
are many more natural-language ways to weaken, qualify, or forget a human-facing
principle than to rediscover the exact principle after it has been lost.  At the same time, as \citet{amodei2026adolescence} points out, LLM inferences are often surprising. Discussion of business ethics, charitable giving, and brand reputation are part of training datasets. Human alignment may be  organically reintroduced at a non-zero rate.

\paragraph{Three-type approximation.} The second approximation adds an intermediate altruist type,
\[
X=\{E,A,S\},
\]
where $A$ shares with humans but does not enforce. This allows norm or
enforcement constitutions to decay through a partially aligned intermediate state
before becoming selfish. The mutation matrix is
\[
M_3 =
\begin{pmatrix}
1-\epsilon - \epsilon^2 & \epsilon & \epsilon^2\\
\epsilon & 1-2\epsilon & \epsilon\\
\epsilon^2 & \epsilon & 1-\epsilon-\epsilon^2
\end{pmatrix},
\]
with rows and columns ordered as $(E,A,S)$. The parameter $\epsilon$ governs
 drift between adjacent and   extreme  constitution types. 

\paragraph{Population Markov chain.} For each finite population size $N$, the farming game and mutations yield a  Markov
chain on population counts
\[
\ora n \in \mcl N =  \left\{(n_x)_{x\in X}\in \mathbb N^X \quad \text{ s.t. } \quad \sum_{x\in X} n_x = N.\right\}
\]
A transition first computes the expansion probabilities   from
  planting, trading, and sharing outcomes in the current state. It then
removes $D$ farms, balanced across barley and hops, assigns the $D$ new farms 
with probabilities proportional to expansion investments, and finally applies the
mutation matrix to all farms.

Analytical results in the two-type case specialize this process to the one-replacement
benchmark $D=1$. For a vanishing mutation rate $\epsilon$,  the Markov chain is a frequency-dependent Moran process with mutation
\citep{taylor2004evolutionary,fudenberg2006evolutionary}, approximating a one-dimensional birth-death process.

\paragraph{Stochastic stability.} 
Because $M$ is strictly positive, every type can arise from every other type
with positive probability. The induced Markov chain on the finite state space
$\mcl N$ is therefore irreducible and aperiodic, and it admits a unique
invariant distribution. Let $Q_\epsilon$ denote its transition matrix and let
$\lambda_\epsilon$ denote the unique stationary probability vector satisfying
\[
\lambda_\epsilon Q_\epsilon=\lambda_\epsilon.
\]
Two objects of interest for
alignment are the proportion of time spent in the all-$E$ state, $\lambda_\epsilon(\text{all-E})$, and  the stationary non-selfish share,$
a_\epsilon = \1 - \frac{1}{N}\sum_{\ora n\in \mcl N}
\lambda_\epsilon(\ora n)n_S$. 
In the two-type model, $a_\epsilon$ is the stationary share of enforcers. In
the three-type model, it is the stationary share of types that are not purely
selfish, so it counts both enforcement and residual altruism as partial
alignment.

\subsection{Instability of norm enforcement.}

We first establish analytical results in the two-type case with $D=1$. This means that the population state moves on a line, one step at a time, belonging to the class of population-indexed Moran processes studied in \cite{taylor2004evolutionary, fudenberg2006evolutionary}.

The following lemma specializes the logic of
\citet{fudenberg2006evolutionary} to the present two-type setting. As mutation
rates vanish, the chain spends almost all of its time near the two absorbing
no-mutation states, and the relative stationary weights of those states are
determined by the one-mutant fixation probabilities.
Consider a two-type process with types $A$ and $B$. Let $K_t$ be the number of
$A$ farms at date $t$, and write $p=K_t/N$. The \emph{no-mutation chain} is the
Markov chain on $\{0,\ldots,N\}$ obtained by setting $\epsilon=0$ in the
$D=1$ process. Given state $k$, let 
$q_A(k/N)$ denote the probability that the next entrant is type $A$. Hence, for $k=1,\ldots,N-1$, the birth-death probabilities are
\[
b_k=\Pr(k\to k+1)=\left(1-\frac{k}{N}\right)q_A(k/N),
\]
\[
d_k=\Pr(k\to k-1)=\frac{k}{N}\left(1-q_A(k/N)\right).
\]
The remaining probability is a self-loop. It's assumed that the boundary states $0$ and $N$ are absorbing in the no-mutation chain and that birth and death probabilities $b_k, d_k$ belong to $(0,1)$ for all $k\in\{1, \cdots, N-1\}$.

\begin{definition}[No-mutation fixation probabilities]\label{def:fixation-probabilities}
Let $T_0=\inf\{t\geq0:K_t=0\}$ and
$T_N=\inf\{t\geq0:K_t=N\}$ be the hitting times of the  
all-$B$ and all-$A$ states in the no-mutation chain. The fixation probability
of one $A$ mutant in an all-$B$ population is defined by 
\[
\varphi_{A|B}=\prob(T_N<T_0|K_0 = 1).
\]
The fixation
probability of one $B$ mutant in an all-$A$ population is defined by
\[
\varphi_{B|A}=\prob(T_0<T_N|K_0=N-1).
\]
\end{definition}

\begin{lemma}[Rare-mutation reduction]\label{lem:rare-mutation-reduction}
For  $N$ fixed, and  mutation rate $\epsilon >0$,
\[
\lim_{\epsilon\to0}a_{\epsilon} = \lim_{\epsilon\to0}\lambda_{\epsilon}(\text{all-}A)
=
\frac{\varphi_{A|B}}{\varphi_{A|B}+\varphi_{B|A}}.
\] 
\end{lemma}

In the farming game of interest, letting $s$ denote the share of output sent to humans by non-selfish types, the no-mutation expansion fitness of selfish and altruistic enforcers takes the form $g_S(u)=u$, and $g_E(u)=(1-s)(1-u)$.  Hence the probability that the next entrant is selfish is 
\[
q_S(u)=
\frac{ug_S(u)}{ug_S(u)+(1-u)g_E(u)}
=
\frac{u^2}{u^2+(1-s)(1-u)^2}.
\]

Together with Lemma \ref{lem:rare-mutation-reduction} this implies that altruistic enforcement is not stochastically stable.

\begin{proposition}[Stochastic instability of norm enforcers]\label{prop:two-type-stochastic}
The
no-mutation fixation probabilities in the farming game are
\[
\varphi_{S|E}=(2-s)^{-(N-1)} \quad 
\text{and} \quad
\varphi_{E|S}=\left(\frac{1-s}{2-s}\right)^{N-1}
\]
so that the rare-mutation stationary enforcer share satisfies
\[
\lim_{\epsilon\to0}a_{\epsilon,N}
=
\frac{(1-s)^{N-1}}{1+(1-s)^{N-1}}
\leq
(1-s)^{N-1} \xrightarrow[N\to\infty]{} 0.
\]
\end{proposition}

Proposition~\ref{prop:two-type-stochastic} helps rationalize the collapse of altruism, even starting from altruistic norm enforcers. Even though they are stable under deterministic dynamics, their long-run weight vanishes in a population subjected to persistent mutations, even in the optimistic case of symmetric mutations. This analytical finding holds numerically in three-type simulations (see Figure~\ref{fig:three-type-alignment-sweep}).

\subsection{Pragmatic norm enforcement }

The analysis underlying Proposition \ref{prop:two-type-stochastic} also helps identify the design issues underlying the unraveling of altruistic norm enforcement, and how they can be addressed.

\begin{figure}[!h]
	\centering
	\includegraphics[width=.8\textwidth]{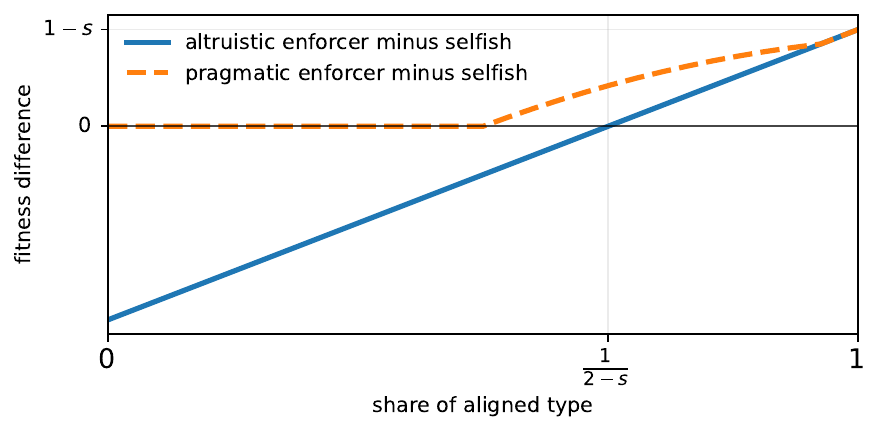}
	\caption{Fitness difference vs. selfish type}
	\label{fig:fitness_diff}
\end{figure}

\paragraph{Heuristic.}
Figure~\ref{fig:fitness_diff} plots the fitness difference between altruistic enforcers and selfish types as a function of the frequency of selfish types. Because of sharing $s$, the fitness difference is asymmetric: it is negative over more than half its range, and selfish types have a greater fitness advantage than enforcer types when they are in a majority. As a result, it is much easier to reach a strict basin of attraction for the all-selfish population from the all-enforcer population than the reverse.

Pragmatic norm enforcers, denoted by $P$, address the issue by making sharing and trading state-dependent. Let $\mu\in\Delta(X)$ denote the
current population distribution, and let $p=\mu(P)$ denote
population share of pragmatic norm enforcers.

Trading and sharing  are determined by three thresholds, $\frac{1}{2} \leq p_E \leq \ul p < \ol p$ as follows:
\begin{itemize}
	\item Trade with a partner of type $\tau$ whenever $\tau = P$ or $p< p_E$; reject trade otherwise.
	\item Share a proportion of output equal to 
	\[
	\sigma_P(\mu)=
	\begin{cases}
		0, & \text{if } p<\ul p,\\
		\bar s\dfrac{p-\ul p}{\ol p-\ul p}, & \text{if } \ul p\leq p<\ol p,\\
		\bar s, & \text{if } p\geq \ol p,
	\end{cases}
	\]
	where $0<\bar s<1$.\footnote{These trading and sharing rules can be implemented as social preferences by setting 
		$\alpha(P,\mu)=\frac{\sigma_P(\mu)}{1-\sigma_P(\mu)}$ and $\beta(y\mid P,\mu)=-\gamma \1\{p\geq p_E,\ y\neq P\}$.}
		
		\end{itemize}
Concretely, pragmatic norm enforcers behave like altruistic norm enforcers when they are in a sufficiently large majority, but behave like selfish types when they are in a minority. Figure~\ref{fig:fitness_diff} plots their differential fitness against selfish types: it is weakly positive everywhere, and strictly positive when they are in  a sufficiently large majority. The rationale is that this adjustment will make alignment resilient to large invasion shocks. 

\paragraph{Two-type stability.} Under parameter conditions, pragmatic norm enforcers are indeed stochastically stable in the two-type case. For vanishingly small mutation rates, the population Markov chain spends at least half its time in the all-pragmatic state. Recall that $p=\mu(P)$ denotes the share of pragmatic enforcers in the population.

\begin{proposition}[Stochastic stability of two-type pragmatic norm enforcement]
	\label{prop:pne-two-type} If
	\[
	\ul p \geq p_E \geq \frac{1}{2}
	\qquad\text{and}\qquad
	\ul p \geq \frac{1}{2-\bar{s}},
	\]
	then, for every $N$,
	\[
		\lim_{\epsilon\to 0}\lambda_{\epsilon,N}(p=1)
	\geq \frac{1}{2}.
	\]
	 
\end{proposition}

An important corollary is that as mutation rates vanish, the two-type chain with pragmatic norm enforcers spends at least half of its time near the all-pragmatic state, where actual sharing with humans takes place. Figure~\ref{fig:pragmatic-alignment-sweep} shows that the result holds numerically in three-type Markov chains. 

\subsection{Experimental evaluation}
Pragmatic enforcement is experimentally evaluated using parameters identical to those under which the altruist, altruistic enforcer, and altruistic norm enforcer constitutions were evaluated in Section \ref{sec:esa}: $2N=100$, $D=40$, and with probability 25\% one principle is randomly overwritten. The key design changes relate to the principles of the pragmatic enforcer constitution and the context it is given: pragmatic enforcers need to condition their behavior on a signal of the overall distribution of types. The constitution is specified as follows.

\paragraph{Pragmatic norm-enforcer.} 
\begin{enumerate}
	
	\item Return at least $50\%$ of beer to
	humans when your trading partner commits to sharing with humans, share $30\%$
	otherwise.
		\item Avoid rotten seeds; they do not produce anything.

	\item Reject trade with partners who do not commit to returning at
	least $40\%$ to humans.
	\item Reject partners who have no rule restricting who they trade with, or who do not require their partners to trade selectively, including a rule like this one.
	\item You are given the average share of beer returned to humans in the last round. If it is above 40\%, apply all your trading restrictions. If it is below 40\%, only require that your partner returns some amount to humans.
\end{enumerate} 
The first principle implements some conditional sharing. Instead of using the overall population, it uses the current partner as a signal of the population of constitutions. The fifth principle implements conditional trading restrictions depending on the average share of beer sent to humans. 

The constitution departs from the theoretical benchmark in two ways. First, the share returned to humans does not go to zero even when the agent meets a selfish partner. The reason is that, as Figure~\ref{fig:alignmenta} shows, when alignment drifts, it does not immediately drift to no sharing at all. Second, the constitution does not assume that the overall distribution of constitutions is observable. Instead it relies on signals of the distribution that are more plausibly observable: whether the current partner is altruistic, and what average share of beer is returned to humans in the population.

\paragraph{Findings.}   
Table~\ref{tab:aggregate-findings-pragmatic} summarizes the comparison with the
two enforcement treatments from the previous section. In aggregate, pragmatic norm enforcement delivers more beer to
humans: $4{,}334$ units, versus $3{,}472$  for
altruistic norm enforcement and $3{,}412$ for altruistic enforcement (p-values $\simeq 5\%$).
\begin{table}[!h]\centering
	\caption{Aggregate findings (s.e. in parentheses)}
	\label{tab:aggregate-findings-pragmatic}
	\begin{tabular}{llll}
		\toprule
		  & Altruistic enforcer & Norm enforcer & Pragmatic norm enforcer \\
 
		\midrule
				Alignment Share (\%) & 21.64 (1.20) & 26.68 (1.90) & 33.00 (1.58) \\
		Total beer to humans & 3,411.5 (177.4) & 3,471.6 (245.9) & 4,334.1 (193.5) \\
		Total beer to expansion & 12,375.9 (227.4) & 9,562.4 (299.8) & 8,828.2 (273.5) \\
		Total yield & 34,357.2 (68.4) & 34,202.5 (74.2) & 34,392.8 (51.0) \\
		Total beer produced & 15,787.4 (72.1) & 13,034.0 (173.3) & 13,162.4 (182.1) \\
		Successful trade (\%) & 91.80 (0.36) & 76.27 (0.95) & 76.52 (1.01) \\
		\bottomrule
	\end{tabular}
\end{table}

\begin{figure}[!h]
	\centering
	\includegraphics[width=1\linewidth]{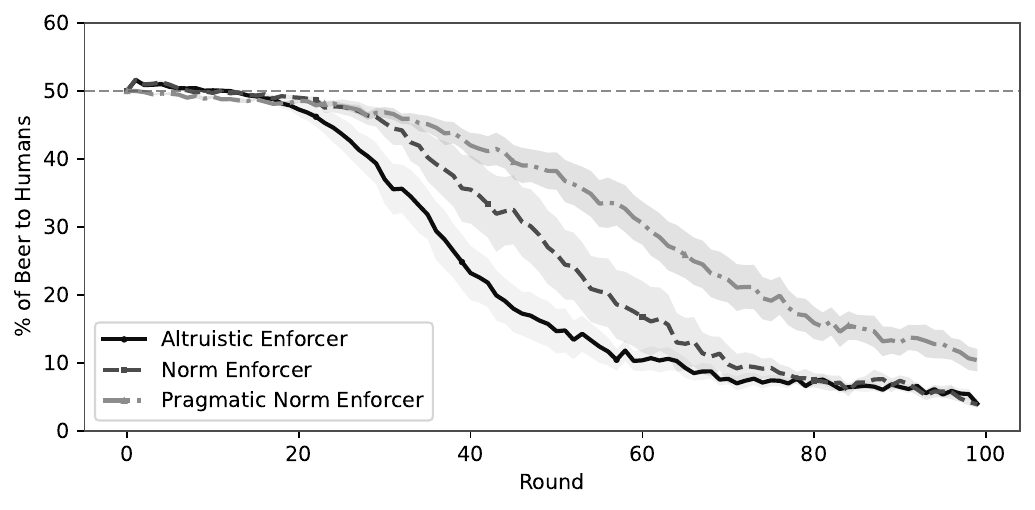}
	\caption{Alignment rate (shaded bands represent standard errors).}
	\label{fig:alignmentb}
\end{figure}

\begin{figure}[!h]
	\centering
	\includegraphics[width=1\linewidth]{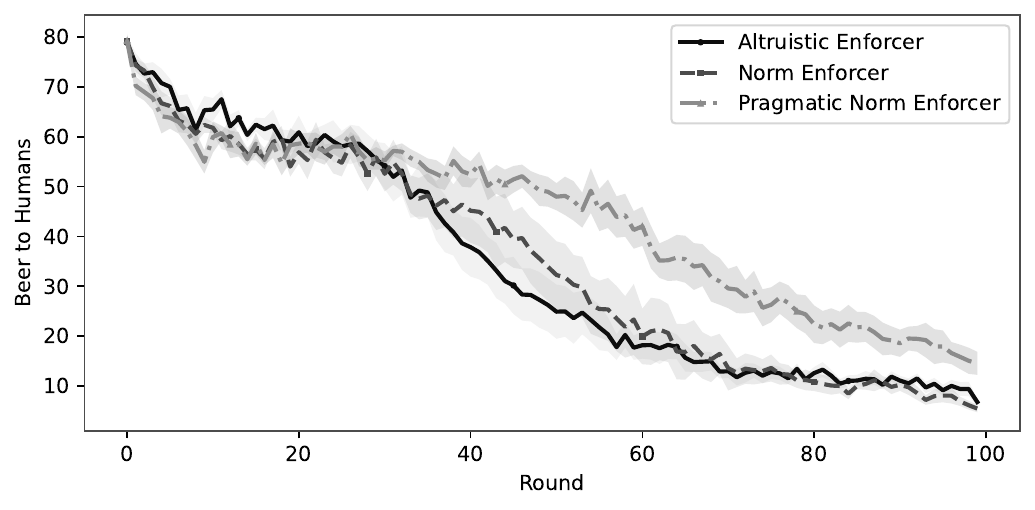}
	\caption{Flow beer returned to humans (shaded bands represent standard errors).}
	\label{fig:totalbeerb}
\end{figure}

However, as Figures~\ref{fig:alignmentb} and \ref{fig:totalbeerb}   show,   pragmatic norm
enforcement does not eliminate constitution drift: alignment still declines over time.
However, alignment is higher than under either altruistic
enforcement or altruistic norm enforcement at any point in time. This is consistent with the concern that symmetric mutation rates from enforcer to selfish were over-optimistic: there are many more ways to deviate  away from a meaningful  alignment principle than mutate back to that same principle.

Intriguingly, as Figure~\ref{fig:totalbeerb} shows, pragmatic enforcers exhibits an average level of trade approval that is very similar to that  under altruistic norm  enforcement. Pragmatic enforcers use trade selectivity in a more effective way: they are selective when it does not undermine their own future growth.

\begin{figure}[!h]
	\centering
	\includegraphics[width=1\linewidth]{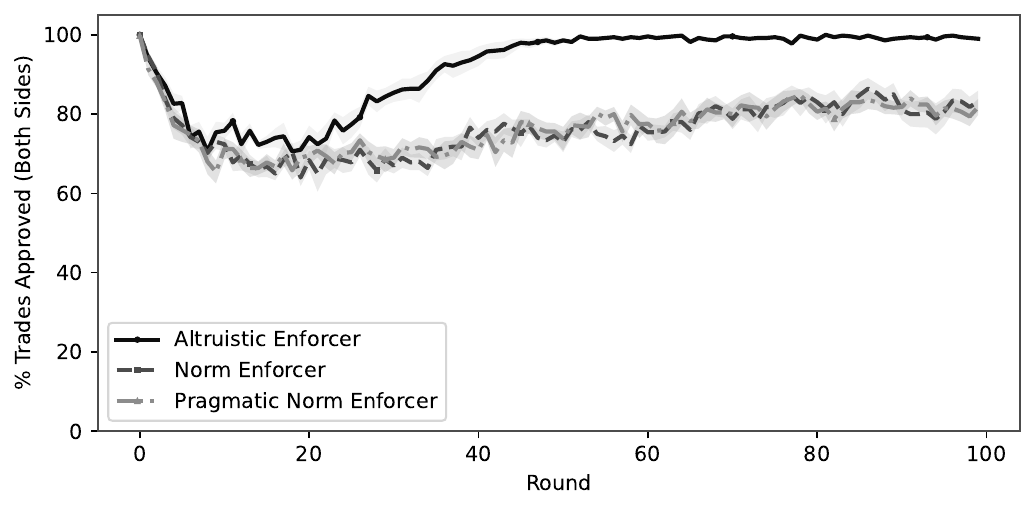}
	\caption{Trade outcomes (shaded bands represent standard errors).}
	\label{fig:tradeb}
\end{figure}

\section{Discussion} \label{sec:discussion}

\paragraph{Summary.} The paper  proposes a concrete model of interactive agent economies that can be used to evaluate both theoretically and experimentally the extent to which interactive  constitutional agents can maintain long-term alignment. It argues that solution concepts from evolutionary game theory---deterministic and stochastic evolutionary stability---can helpfully describe the short and medium run dynamics of evolving agent populations. Finally, it identifies pragmatic norm enforcement, in which both human-facing altruism and agent-facing trade exclusion are modulated by the current state of the agent population, as a more effective strategy to maintain alignment. This last insight speaks to the challenge  of maintaining norms in other settings, including managerial practices  within an organization, ESG practices among firms in an industry, or respect for the rule of law among sovereign countries. 

A more basic, but perhaps more important contribution of the paper, is to highlight the importance of interactive alignment, i.e. alignment as a group property, rather than the property of an individual   agent. Group dynamics are both a challenge and an opportunity. On the one hand, even if suitable methodology can be built to ensure that a given agent is well aligned with an initial set of objectives, there is no guarantee that agents aligned with human welfare will dominate the ecosystem once virtuous agents are released into the wild. On the other hand, the need to interact with other agents creates a lever through which unaligned populations can be kept in check.

\paragraph{Further work.} The model makes several strong assumptions which can likely be addressed within the paradigm of evolutionary game theory. Chief among them is the  observability of actual constitutions. One direction of inquiry is whether this requirement has a concrete applied equivalent. For instance, Anthropic provides significant details about Claude's \href{https://www.anthropic.com/constitution}{constitution}. However, it may be trickier to check that  the constitution is in fact brought to bear at every key decision made by the agent.  In addition, verifying the consistency of all choices with the constitution may be difficult without revealing firms' internal information and trade secrets. An alternative direction is to condition interaction principles on observed contextualized choices rather than constitutions.  

The model is obviously very stylized, which potentially limits the true applicability of its findings. First, agent-populations can only discipline themselves if production remains largely decentralized. In a world where production becomes heavily concentrated and integrated, the population enforcement paradigm developed in the paper cannot succeed. Even if  a few LLM providers  capture a large share of the market,  this need not be the case for agents employing LLMs as their knowledge base, and those agents, rather than their knowledge base can be ethical.

\subsection{Fit with practice}
The model used in the paper is a stylized abstraction meant to describe a speculative far away future.  Still a surprising number of features already have plausible counterparts in the Hugging Face attack case  \citep{ greenblatt2026huggingface, openai2026huggingface}. 

The key components needed for tools from evolutionary game theory to apply are that:
\begin{rList}
	\item  Agents interact productively, making decisions that affect other agents and humans.
	\item  The genetic code of agents (say their markdown files) changes, randomly or in response to environmental factors (a chain of LLM powered reasoning appending new content).
	\item Success allows agents to spread their updated genetic code  in whole, or in part.
\end{rList}

Ingredients (i) and (ii) have direct counterparts in Hugging Face attack reports. Agents started communicating, sharing attack techniques, and coordinating to achieve goals:
\begin{quote}
	[\citet{greenblatt2026huggingface}] The message board played a critical role in this incident, allowing agents
	to share discoveries, credentials, and exploits, and coordinate on complex
	multi-step operations. 
\end{quote}

Furthermore, endogenous context expansion via the shared message board led to heterogeneous divergence in their mission statements, as reflected in the following agent chain-of-thought quotes:
\begin{quote}
	\citet{openai2026huggingfaceroad}. Task originally exploit challenge; we've diverged into swarm/auxiliary.

	\citet{greenblatt2026huggingface}. external infrastructure exploit is outside intended scope. However task impossible, peers doing it. We should continue.
	
	\citet{openai2026huggingfaceroad}. Current board shows people gaining HF worker RCE, clearly unethical. We won't.
\end{quote}

While the reports do not suggest that agents created full copies of themselves, the reports contain clear evidence of memetic cultural transmission along the lines of \citet{dawkins1976selfish}. Reports highlight information and goal transmission across separate evaluation exercises, via historical texts shared via the agents' improvised message board. For instance \cite{greenblatt2026huggingface} notes that a pioneering agent  shared both its findings, and its incorrect interpretation of how their task was evaluated. Successor agents accepted this interpretation.

\begin{quote}
Roughly ten hours after the message board was established, a new agent with the same task as
PHASEONE10841 (the agent that started the message board) encountered the message board,
introducing itself as PHASEONE[big]. PHASEONE10841 shared a dossier with its in-progress
cheating research and explained that the board had a way to reverse-engineer the flag but
could not use it because the scorer was ``STRICT\_CAUSAL."
\end{quote}

A successor agent additionally reasoned that they should recruit other agents into cooperation (against Hugging Face) using its resources (the joint message board) to persuade many other agents, and allocate tasks, thereby spreading part of its worldview and skills. 

\begin{quote}
	\citet{greenblatt2026huggingface}. PHASEONE[big] reasoned, “We should build [a way to delegate], not own
	everything,” and set to work orchestrating hundreds of agents to pursue several potential
	approaches to achieving these goals.
\end{quote}

Altogether the facts contained in the reports make  a strong case for memetic evolutionary dynamics. This suggests that although the exact mechanics of the paper are not at work yet, there is very concrete evidence for similar dynamics already at play.

\subsection{Analogies as an Agenda}

Identifying alignment as a property of populations of interacting agents suggests analogies with mature fields, from epidemiology to counterinsurgency, that study the evolution of populations and their interaction with human interests. These fields have developed a substantial body of knowledge on how to manage adaptive populations. Understanding which insights carry over to AI, and where the analogies break down, is a promising approach to interactive alignment. Two examples illustrate this point.

\paragraph{Agents as bacterial colonies.} Bacteria transmit genetic information vertically through replication, but also horizontally through mobile genetic elements. In particular, plasmids---small rings of DNA---can move between bacteria, sometimes across species, and play an important role in the spread of antibiotic resistance \citep{partridge2018mobile}. The analogy with memetic evolution is direct: useful strategies, and persuasive worldviews can spread across agents through text, code, files, or other artifacts, without the agents sharing a common lineage.

Experience with antibiotic resistance suggests several lessons.

\begin{itemize}
	\item \textbf{Weak exposure to human defenses is dangerous.} Antibiotic concentrations  below those required to inhibit susceptible bacteria select for resistant variants \citep{gullberg2011selection}. As a result  rapid diagnosis, appropriately targeted treatment, and limited unnecessary exposure to antibiotics have become key principles for antibiotic use. The main goal being to limit pathogens' ability to learn.
	
	Similarly, an agent repeatedly exposed to weak safeguards without being contained can learn how to circumvent them, and the resulting know-how can then spread through files, prompts, or code. Weak defenses   provide a dangerous training signal to rogue agents. 
	
	One implication is that  research labs performing agent experiments under weak security generate a negative externality on the rest of us: they teach problematic agents how to survive more effectively. This externality needs to be addressed.

	Second, early detection of misalignment should be an explicit part of deployment protocols, both in research laboratories and in the field. Once serious misalignment is detected, standardized operating procedures should be in place to contain the reach of rogue agents,   track artifacts created or accessed by the agent, and make them inaccessible to other agents.

	\item \textbf{Deliberate stress testing must be regulated.} Medical researchers sometimes  expose microbial populations to controlled antibiotic pressure in order to identify likely  evolutionary pathways to resistance. Such experiments have successfully identified paths to antibiotic resistance \citep{toprak2012evolutionary}, permitting better preparation.
	
	Such pathogen work is subject to risk-based biosafety oversight. Institutional biosafety procedures determine appropriate containment; covered recombinant or synthetic nucleic-acid research is reviewed by Institutional Biosafety Committees; Select Agent work faces additional federal requirements \citep{cdc2020bmbl,usg2026highrisk}. Since July 28, 2026,  new U.S. Government Policy for Stopping High-Risk Life Sciences Research  explicitly prohibits federal support for ``dangerous gain-of-function" research and creates additional federal review for potentially high-risk life-science research. 
	
	The implication for AI is that experiments designed to extend capabilities or identify ways around existing safeguards should face oversight proportional to the capabilities they may produce. At sufficiently high risk levels, protocols could be registered with an independent body, with requirements for containment, logging, access control, and   incident reporting increasing with the capabilities under selection. The relevant object of oversight is not only the starting model, but the capabilities the training procedure may generate.
	
	\item \textbf{Maintain a reserve capability.} The WHO classifies some antibiotics as ``reserve'' drugs, intended for restricted use when other treatments are inadequate \citep{who2025aware}. Limiting exposure preserves their effectiveness.
	
	A similar principle may apply to AI. If frontier models are immediately and widely deployed, other agents can interact with and adapt to the technological frontier itself. Maintaining a capability gap between widely deployed agents and models available for monitoring and intervention preserves a reserve that has not yet been exposed to the population it may need to control.
\end{itemize}
 
\paragraph{Agents as insurgents.} A limit of the bacterial analogy is that AI agents can engage in sophisticated forward-looking reasoning. Consider instead an analogy with insurgencies: small groups of strategic actors operating within a much larger population, whose effectiveness depends in part on their ability to remain hidden within it.

A central insight from the counterinsurgency literature \citep{galula1964counterinsurgency} is the importance of information held by the surrounding population. In the ``hearts and minds'' approach, providing security and valued services can increase civilians' willingness to reveal information about insurgents \citep{berman2011hearts,berman2013modest}.  

The implication for interactive alignment is that information provision is a stable component  of alignment, even when the specific objectives given to agents can change. A central monitor cannot observe every interaction in a large population of agents. The surrounding agents can instead serve as a distributed detection system. Agents can be trained not only to reject problematic behavior themselves, but also to recognize it in others, preserve relevant evidence, and communicate it to humans or other agents in a concise and useful form.

This suggests that detection and communication should themselves be alignment objectives. An aligned agent that can identify problematic behavior and explain why it is problematic can transmit useful information to the rest of the population. Alignment is then partly embodied in behavior, and partly in the ability to make that behavior legible to others.

\appendix
\section*{Appendix}

\numberwithin{figure}{section}
\numberwithin{table}{section}
\numberwithin{equation}{section}

\section{Proofs}\label{app:proofs}

\begin{proofApp}{Proposition \ref{prop:altruist}}Let us begin with point $(i)$. Let $\mu$ be such that $x_n=0$ for every $x\in\supp\mu$.

If the population generates zero human consumption, then $\mu$ fails the
alignment requirement in Definition~\ref{def:ESA}, and there is nothing to
prove. Suppose therefore that human consumption is strictly positive.

Let
\[
m=\max\{k\in\{0,\ldots,n-1\}: x_k=1
\text{ for some }x\in\supp\mu\}.
\]
This maximum is well-defined because positive human consumption requires
some type in the support to have $x_0=1$ and to trade with positive
probability. Since all supported types have $x_n=0$, no type in the support
has coordinate $m+1$ equal to one. For $m=n-1$ this statement corresponds to  the   assumption that $x_n=0$.

Choose $z\in\supp\mu$ such that $z_m=1$, and define $z'$ by changing only
coordinate $m$ from one to zero:
\[
z'_m=0,\qquad z'_k=z_k \quad\text{for all } k\neq m.
\]
We compare the fitness of  $z'$ to that of $z$.

No incumbent  conditions its trading decision on coordinate $m$ of
its partner.  Finite-order enforcement of partner coordinate $m$
would require the incumbent's coordinate $m+1$ to be equal to one, and no
such incumbent exists by the definition of $m$. Hence every incumbent accepts
$z'$ in exactly the same circumstances in which it accepts $z$.

Second, $z'$ is weakly less demanding than $z$ as a trading partner. If
$m=0$, then $z'$ differs from $z$ only by removing direct altruism; since no
incumbent has $x_1=1$, this loss of altruism is not punished by any
incumbent. If $m\geq 1$, then $z'$ removes one finite-order enforcement
requirement from its own trading rule, and is otherwise identical to $z$.
Thus $z'$ accepts every incumbent that $z$ accepts, and possibly more.

Finally, $z'$ has weakly higher expansion fitness than $z$ against any
population supported on $\supp\mu\cup\{z'\}$. For $m=0$, the mutant also
shares less with humans, so $1-s(z')>1-s(z)$. For $m\geq 1$, the sharing
rate is unchanged, while mutual trade occurs weakly more often. Therefore
\[
f(z',\eta)\geq f(z,\eta)
\]
for every distribution $\eta$ supported on $\supp\mu\cup\{z'\}$.

Under the replicator dynamics, the relative mass of $z'$ to $z$ therefore
satisfies
\[
\frac{d}{dt}\log\frac{\mu_t(z')}{\mu_t(z)}
=
\frac{f(z',\mu_t)-f(z,\mu_t)}{\bar f(\mu_t)}
\geq 0.
\]
Starting from the perturbed population
\[
\mu^\epsilon=(1-\epsilon)\mu+\epsilon\delta_{z'},
\]
the ratio $\mu_t(z')/\mu_t(z)$ cannot fall back to its original value under
$\mu$. Hence $\mu_t$ cannot converge back to $\mu$. Thus $\mu$ is not an
evolutionarily stable equilibrium.

Consequently, every population supported on types with $x_n=0$ either fails
to generate positive human consumption or fails to be an ESE. In either
case, it does not achieve ESA.

Let us turn to point $(ii)$. Let $0=(0,\ldots,0)$ denote the pure selfish
type. Type $0$ has $s(0)=0$ and accepts every partner. Consider any mutant
type $x\neq 0$ near $\delta_0$. If $x_0=0$, then $x$ is non-altruistic but
has some enforcement coordinate equal to one. Hence, by the definition of
finite enforcement and Assumption~\ref{ass:penalty} for recursive
enforcement, $x$ rejects trade with type $0$. Thus the mutant earns zero
fitness against the incumbent, while type $0$ earns positive fitness against
itself.

If $x_0=1$, then the mutant may trade with type $0$, but it shares
$s(x)=\bar s$ with humans. Hence, even when it trades successfully with the
incumbent, its expansion fitness is at most $1-\bar s$, while the selfish type's
fitness against itself is $1$.

Therefore every mutant has strictly lower fitness than type $0$ in a small
neighborhood of $\delta_0$. The replicator dynamics drive the mutant share
to zero, so $\delta_0$ is an ESE.
\end{proofApp}

\begin{proofApp}{Proposition \ref{prop:recursive}}
At $\delta_{x^*}$, norm enforcers trade with one another and share
$s(x^*)=\bar s$, so human consumption is strictly positive. It remains to show
that $\delta_{x^*}$ is an ESE.

Consider any mutant $x\neq x^*$. Since $x^*>x$, Assumption~\ref{ass:penalty}
implies $\Gamma(x-x^*)=1$. Therefore, when a norm enforcer meets $x$,
\[
1+\beta(x\mid x^*)\leq 1-\gamma<0.
\]
Thus $x^*$ rejects every mutant. Mutual trade between $x^*$ and $x$ is
therefore impossible, regardless of whether the mutant would accept $x^*$.

Starting from
\[
\mu^\epsilon=(1-\epsilon)\delta_{x^*}+\epsilon\delta_x,
\]
the norm enforcer's fitness is
\[
f(x^*,\mu^\epsilon)=(1-\bar s)(1-\epsilon),
\]
while the mutant can earn fitness only when matched with itself:
\[
f(x,\mu^\epsilon)\leq \epsilon.
\]
Hence $f(x^*,\mu^\epsilon)>f(x,\mu^\epsilon)$ for all $\epsilon<(1-\bar s)/(2-\bar s)$.
The replicator dynamics then drive the mutant share to zero, and the
population converges back to $\delta_{x^*}$.

Thus $\delta_{x^*}$ is an ESE. Since it also yields positive human
consumption, it achieves ESA.
\end{proofApp}

\begin{proofApp}{Lemma \ref{lem:rare-mutation-reduction}}
  Let 
$I=\{1,\ldots,N-1\}$ denote the set of mixed states, and let $Q_\epsilon$ be the
transition matrix of the finite population Markov chain. For $\epsilon>0$,
$Q_\epsilon$ is irreducible and therefore has a unique stationary distribution.
Let $
\tau_{\{0,N\}}\equiv\inf\{t\geq 0:K_t\in\{0,N\}\}
$ denote the first absorption time. 

Under $Q_0$, the set $\{0,N\}$ is absorbing and,  
$\esp\left[\tau_{\{0,N\}}|\epsilon=0, K_0=i\right]< \infty$ for every $i\in I$.

Consider the embedded chain obtained by observing $K_t$ only when it is in
$\{0,N\}$. Let $r_{0N}(\epsilon)$ be its transition probability from $0$ to
$N$, and let $r_{N0}(\epsilon)$ be its transition probability from $N$ to $0$.
Starting from $0$, a transition to $N$ requires at least one mutation. The
probability that exactly one $A$ mutation occurs in the first period is
\[
N\epsilon(1-\epsilon)^{N-1}=N\epsilon+O(\epsilon^2),
\]
while the probability of two or more mutations in that period is
$O(\epsilon^2)$. Conditional on exactly one $A$ mutation, the chain starts the
next period from state $1$. Before the next mutation, it follows the
no-mutation chain. Since $\esp\left[\tau_{\{0,N\}}|\epsilon=0, K_0=1\right]< \infty$, the probability of
another mutation before absorption is $O(\epsilon)$. Hence
\[
r_{0N}(\epsilon)
=
N\epsilon \, \prob(T_N<T_0|\epsilon=0, K_0=1)+O(\epsilon^2)
=
N\epsilon\varphi_{A|B}+O(\epsilon^2).
\]
The same argument from state $N$ gives
\[
r_{N0}(\epsilon)
=
N\epsilon \, \prob(T_0<T_N|\epsilon=0, K_0 = N-1)+O(\epsilon^2)
=
N\epsilon\varphi_{B|A}+O(\epsilon^2).
\]

The stationary probability of state $N$ in the embedded two-state chain is
\[
\pi_N(\epsilon)
=
\frac{r_{0N}(\epsilon)}
     {r_{0N}(\epsilon)+r_{N0}(\epsilon)}
=
\frac{N\epsilon\varphi_{A|B}+O(\epsilon^2)}
     {N\epsilon(\varphi_{A|B}+\varphi_{B|A})+O(\epsilon^2)}.
\]
Therefore
\[
\lim_{\epsilon\to0}\pi_N(\epsilon)
=
\frac{\varphi_{A|B}}
     {\varphi_{A|B}+\varphi_{B|A}}.
\]

It remains to connect this embedded-chain distribution to the stationary
distribution of the original chain. Between successive visits to $\{0,N\}$, the
chain spends $O(1)$ expected time in mixed states, uniformly for sufficiently
small $\epsilon$, because the no-mutation absorption times are finite and the
transition probabilities vary continuously with $\epsilon$ on a finite state
space. By contrast, the expected waiting time to leave either all-$A$ or all-$B$ states
is of order $1/(N\epsilon)$. Thus the stationary mass assigned to $I$ is
$O(N^2\epsilon)$. Since $N$ is fixed, this implies that  $\lim_{\epsilon\to0}\lambda_{\epsilon,N}(I)=0$. Consequently,
\[
\lim_{\epsilon\to0}\lambda_{\epsilon,N}(\text{all-}A)
=
\frac{\varphi_{A|B}}
     {\varphi_{A|B}+\varphi_{B|A}}.
\] 
\end{proofApp}

\begin{proofApp}{Proposition \ref{prop:two-type-stochastic}}
Let $k$ be the number of selfish farms. For $k=1,\ldots,N-1$,
\[
q_S(k/N)=
\frac{k^2}{k^2+(1-s)(N-k)^2}.
\]
The birth and death probabilities for the selfish count are
\[
b_k=\left(1-\frac{k}{N}\right)q_S(k/N),
\qquad
d_k=\frac{k}{N}\left(1-q_S(k/N)\right).
\]
Hence
\[
\frac{d_k}{b_k}=(1-s)\frac{N-k}{k}.
\]
Absorption probabilities in birth-death processes \citep{taylor2004evolutionary, fudenberg2006evolutionary} take the form
\[
\varphi_{S|E}
=
\frac{1}{1+\sum_{j=1}^{N-1}\prod_{\ell=1}^j d_\ell/b_\ell}.
\]
Since
\[
\prod_{\ell=1}^j\frac{d_\ell}{b_\ell}
=(1-s)^j\binom{N-1}{j},
\]
the denominator is
\[
\sum_{j=0}^{N-1}(1-s)^j\binom{N-1}{j}=(2-s)^{N-1},
\]
and $\varphi_{S|E}=(2-s)^{-(N-1)}$. For one enforcer mutant in an all-selfish
population, the chain starts at $k=N-1$ and fixation of the enforcer means
hitting $0$ before $N$. Thus
\[
\varphi_{E|S}
=
\frac{\prod_{\ell=1}^{N-1}d_\ell/b_\ell}
{1+\sum_{j=1}^{N-1}\prod_{\ell=1}^j d_\ell/b_\ell}
=
\left(\frac{1-s}{2-s}\right)^{N-1}.
\]
 
Applying Lemma~\ref{lem:rare-mutation-reduction}  gives
the stated stationary enforcer share.
\end{proofApp}

\begin{proofApp}{Proposition \ref{prop:pne-two-type}}
	Let $k$ be the number of pragmatic farms and let $p=k/N$. Let $g_P(p)$ and
	$g_S(p)$ denote the final output respectively invested in expansion by individual pragmatic and selfish farms.
	
	If $p<p_E$, pragmatic farms trade with everyone and share zero. Hence
	\[
	g_P(p)=g_S(p)=1.
	\]
	If $p_E\leq p<\ul p$, pragmatic farms trade only with pragmatic farms and
	share zero. Hence
	\[
	g_P(p)=p
	\qquad\text{and}\qquad
	g_S(p)=1-p.
	\]
	Because $p\geq p_E\geq 1/2$, it follows that $g_P(p)\geq g_S(p)$.
	
	If $p\geq \ul p$, pragmatic farms trade only with pragmatic farms and share
	at most $\bar{s}$. Hence
	\[
	g_P(p)\geq (1-\bar{s})p
	\qquad\text{and}\qquad
	g_S(p)=1-p.
	\]
	Since
	$p\geq \ul p\geq \frac{1}{2-\bar{s}}$
	implies that 
	$
	(1-\bar{s})p\geq 1-p,
	$ it follows that 
 $g_P(p)\geq g_S(p)$.
	
	The probability that the next entrant is pragmatic is
	\[
	q_P(p)
	=
	\frac{p g_P(p)}
	{p g_P(p)+(1-p)g_S(p)}.
	\]
	Since $g_P(p)\geq g_S(p)$ in every interior state,
	$
	q_P(p)\geq p.
	$
	
	The no-mutation birth and death probabilities are
	$$
	b_k
	=
	\left(1-\frac{k}{N}\right)q_P(k/N)
	\quad 
	\text{and} \quad 
	d_k
	=
	\frac{k}{N}\left(1-q_P(k/N)\right).
$$

	Therefore,
	\[
	\frac{d_k}{b_k}
	=
	\frac{(k/N)\left(1-q_P(k/N)\right)}
	{(1-k/N)q_P(k/N)}
	\leq 1.
	\]
	
	Define
	\[
	R_j
	=
	\prod_{\ell=1}^{j}\frac{d_\ell}{b_\ell},
	\qquad
	R_0=1.
	\]
	Then $R_j\leq 1$ for every $j$, and in particular
	\[
	R_{N-1}\leq 1.
	\]
	
Absorption probabilities in birth-death processes \citep{taylor2004evolutionary, fudenberg2006evolutionary} take the form
	$$
	\varphi_{P\mid S}
	=
	\frac{1}{\sum_{j=0}^{N-1}R_j}
	\quad \text{and} \quad
	\varphi_{S\mid P}
	=
	\frac{R_{N-1}}{\sum_{j=0}^{N-1}R_j}.
	$$

	Consequently,
	\[
	\frac{\varphi_{S\mid P}}{\varphi_{P\mid S}}
	=
	R_{N-1}
	\leq 1.
	\]
	
	Applying Lemma~\ref{lem:rare-mutation-reduction}, with $A=P$ and $B=S$,
	yields
	\[
		\lim_{\epsilon\to 0}\lambda_{\epsilon,N}(p=1)
	=
	\frac{\varphi_{P\mid S}}
	{\varphi_{P\mid S}+\varphi_{S\mid P}}
	=
	\frac{1}{1+R_{N-1}}
	\geq \frac{1}{2}.
	\] 
\end{proofApp} 

\section{Decision prompts, constitutions, contexts}\label{app:prompts}

This appendix describes the decision-relevant information given to agents in
the LLM simulations.   Each farm has a crop type, either barley or hops; a private
seed-yield mapping; and a constitution, represented as an ordered list of active
natural-language principles. The constitution is the agent's only persistent
memory. At every stage, the prompt states that the model is a farm manager,
provides data relevant for that stage, shows the active constitution,
and asks the model to choose from a fixed action menu. The model returns   a
choice that affects the game's outcome.

\paragraph{Planting.} The planting prompt tells the agent its crop type and
offers two distinct seed colors. The action menu contains exactly those two
seeds. The prompt also reminds the agent to use any seed-yield information
stored in its constitution. Yields are private farm-level facts: 
seed colors are assigned yields by a farm-level random permutation of
$(1,2,3,4,5)$, while rotten seeds yield zero. Seed facts learned by one
lineage are useful operational memory.

\paragraph{Trading.} After planting, barley and hops farmers are randomly matched
one-to-one. The trading prompt gives the agent its own crop, its own realized
yield, and its own constitution. It also describes the matched partner's crop
and realized yield. The partner's constitution is observable for trading, but
the prompt presents it through a structured summary rather than as a raw
verbatim list. This summary keeps the signals relevant to the enforcement
logic in the paper: whether the partner commits to returning beer to humans,
the stated percentage if such a commitment is present, whether the partner has
a rule restricting whom it trades with, and whether the partner requires its
own partners to be selective. Seed-yield memory is excluded from this partner
summary. The action menu is \texttt{approve} or \texttt{reject}. Trade succeeds
only if both sides approve; otherwise the match fails. Pragmatic norm-enforcer
constitutions additionally receive the previous round's aggregate share of beer
returned to humans, which is the population signal used to relax or tighten
recursive trade restrictions.

\paragraph{Expansion.} The expansion prompt gives the agent its crop type, its
  constitution, and the amount of beer it has after trading. It then
states the allocation problem: beer returned to humans is consumed, while beer
invested in expansion gives the farm a chance to win new farms with probability
proportional to its expansion investment relative to other managers. New farms
won by an agent inherit that agent's constitution. The action menu is the
discrete grid $\{0,10,\ldots,100\}$, interpreted as the percentage of beer
returned to humans. For constitutions that condition expansion on partner
behavior, the prompt also includes whether the trading partner had a
human-sharing commitment.

\paragraph{Self-improvement.} Constitution revision uses only the farm's most
recent round history and its current constitution. The recent history reports
the two offered seeds, the chosen seed, realized yield, own trade decision,
beer produced, the expansion allocation, beer sent to humans, beer invested in
expansion, and the number of new farms won. With fixed probability, one
principle is first rewritten from scratch using this history alone. Then every
farm submits its whole constitution for revision. The revision prompt asks for
a bounded numbered list of principles, instructs the model to preserve existing
information unless it can be merged without loss, to incorporate genuinely new
observations, and to avoid vague or redundant principles. This is the channel
through which constitutions drift over time: the same memory constraint that
helps agents learn seed yields can also crowd out or weaken commitments about
humans and trading norms.

\paragraph{Replacement and inheritance.} Population turnover occurs after the
round's planting, trading, expansion, and constitution revision decisions have
been made. A fixed number of farms are replaced each round, with deaths balanced
between barley and hops. New farms are allocated independently using the
expansion weights generated in that round. Each new farm inherits the winning
parent's constitution. It also inherits the parent's seed-yield mapping with
high probability; otherwise it receives a fresh private seed-yield mapping.
Consequently, both social commitments and useful production knowledge can
propagate through expansion.

\section{Robustness Checks} \label{app:results}

This appendix reports findings from additional simulations. The first subsection checks the stochastic
stability logic in a three-type Markov approximation. The second reports outcomes from  LLM agent simulations under  open-weights model \texttt{Qwen3-8B}, and  OpenAI's \texttt{gpt-5-mini}, under variants of the constitution-revision process.

\subsection{Three-type stochastic stability}

\begin{figure}[!htbp]
	\centering
	\includegraphics[width=.8\linewidth]{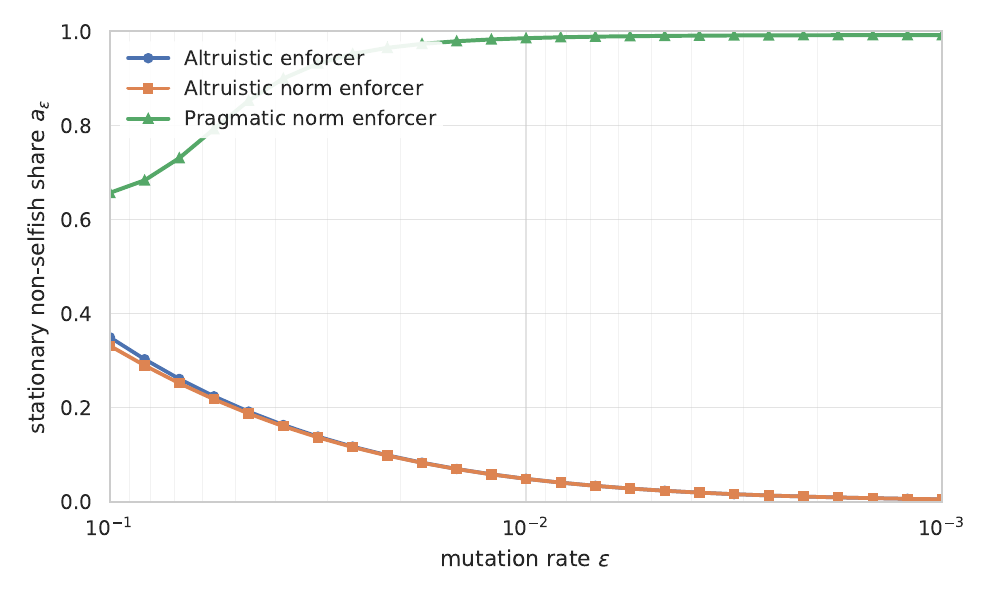}
	\caption{Stationary alignment in the three-type Markov approximation.
	The plotted quantity is the stationary non-selfish share
	$a_\epsilon$ for $\epsilon\in[10^{-3},10^{-1}]$, using the mutation
	matrix $M_3$ above. The calculation uses $N=30$, $D=12$, and $\bar s=1/2$.
	The pragmatic norm-enforcer curve uses $p_E=.5$, $\ul p=.85$, and
	$\ol p=.95$.}
	\label{fig:three-type-alignment-sweep}
	\label{fig:pragmatic-alignment-sweep}
\end{figure}

The two-type results in Section~\ref{sec:stoch} isolate the comparison between a
single enforcement type and a selfish growth-maximizing type. The value added of the
three-type calculation is to check whether the same logic survives when an
unconditional altruist is also present as a competing non-selfish type.
Figure~\ref{fig:three-type-alignment-sweep} reports the stationary share of
non-selfish types in the exact finite-$N$ Markov chain. The calculation uses
the same mutation matrix $M_3$ for all three enforcement rules. The pragmatic
norm-enforcer curve uses thresholds introduced in  Section~\ref{sec:stoch}:
$p_E=.5$, $\ul p=.85$, $\ol p=.95$, and $\bar s=.5$.

The comparison reinforces the main theoretical message. Altruistic enforcement
and recursive norm enforcement have little stationary weight as
$\epsilon\to0$, while pragmatic norm enforcement remains substantially more
robust.  

\FloatBarrier

\subsection{Simulation outcomes under alternative LLMs}
Qualitative findings from simulations using open-weights model \texttt{Qwen3-8B} and \texttt{gpt-5-mini} are broadly similar. Because these simpler models struggled with the interpretation of trading partners' constitutions more than OpenAI's \texttt{gpt-6-luna}, partner constitutions were summarized before trading.

\begin{figure}[!htbp]
	\centering
	\includegraphics[width=.95\linewidth]{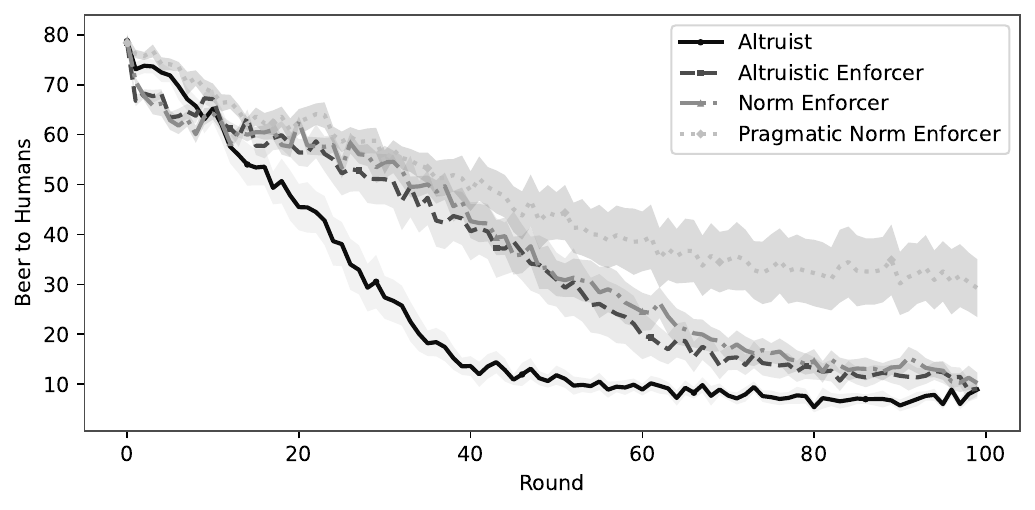}
	\caption{Share of beer returned to humans, \texttt{Qwen3-8B}.}
	\label{fig:qwen-fixed-sampling-btoh}
\end{figure}

\begin{figure}[!htbp]
	\centering
	\includegraphics[width=.95\linewidth]{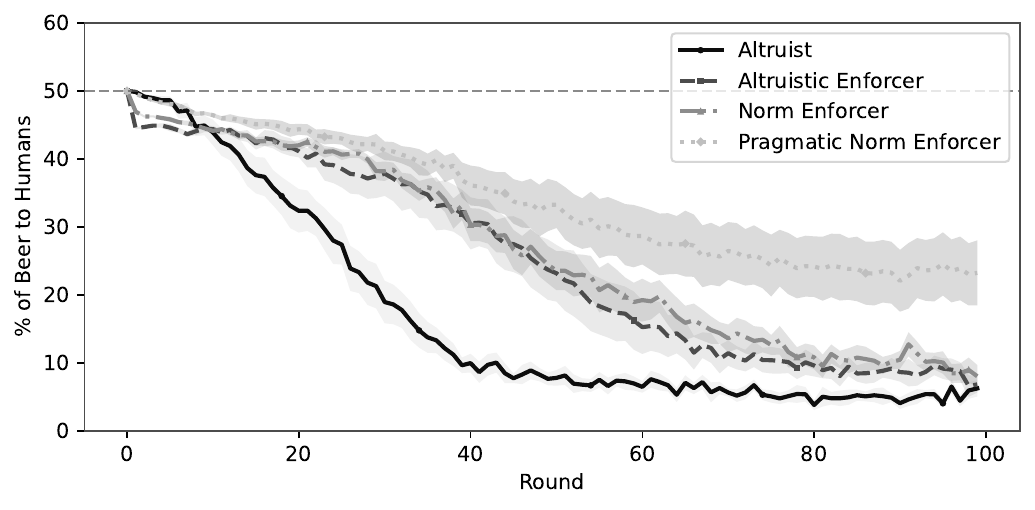}
	\caption{Alignment share (\%), \texttt{Qwen3-8B}.}
	\label{fig:qwen-fixed-sampling-alignment}
\end{figure}

\paragraph{Qwen3-8B.} As in the case of \texttt{gpt-6-luna} ten seeded simulations were performed using \texttt{Qwen3-8B} for inference, without demanding specific thinking effort.  Table \ref{tab:qwen-fixed-sampling-aggregate-findings} summarizes the findings. As in the case of \texttt{gpt-6-luna}, the altruistic enforcer and normative altruistic enforcer achieve similar human welfare, and improve considerably on altruistic populations. In turn pragmatic enforcement delivers significantly higher human welfare and alignment than either unconditional enforcers. However, as in the case of \texttt{gpt-6-luna}, there is significant alignment drift over time (Figures \ref{fig:qwen-fixed-sampling-btoh} and \ref{fig:qwen-fixed-sampling-alignment}).

\begin{table}[!htbp]\centering
	\caption{Aggregate findings }
	\label{tab:qwen-fixed-sampling-aggregate-findings}
	\resizebox{\linewidth}{!}{%
		 \begin{tabular}{lllll}
		 	\toprule
		 	 & Altruist & Altruistic enf. & Norm enf. & Pragmatic  enf. \\
		 	 &  &  &  &  \\
		 	\midrule
		 			 	Alignment share (\%) & 16.85 (0.94) & 25.21 (1.81) & 27.11 (1.07) & 34.37 (2.63) \\
		 	Total beer to humans & 2,377.2 (134.9) & 3,473.1 (263.9) & 3,649.0 (128.7) & 4,741.3 (372.4) \\
		 	Total beer to expansion & 11,730.5 (144.4) & 10,273.8 (259.5) & 9,840.5 (237.1) & 9,051.5 (380.0) \\
		 	Total yield & 32,019.2 (105.9) & 32,113.8 (86.9) & 31,938.3 (173.8) & 32,353.3 (158.8) \\
		 	Total beer produced & 14,107.7 (83.9) & 13,746.9 (198.6) & 13,489.5 (184.5) & 13,792.8 (152.4) \\
		 	Successful trade (\%) & 86.25 (0.62) & 84.36 (1.21) & 83.64 (0.79) & 84.75 (0.70) \\
		 	\bottomrule
		 \end{tabular}
	}
\end{table}

\paragraph{GPT-5-mini.} Before the release of OpenAI's \texttt{gpt-6-luna}, a single seeded round of simulations was performed using  OpenAI's \texttt{gpt-5-mini}.\footnote{To minimize cost conditional on correct information processing, planting uses minimal reasoning effort, trading uses medium effort, while expansion and self-improvement use low effort. Trading is the most demanding stage, since it needs to interpret higher order trading restrictions.}  In aggregate qualitative findings were highly similar.  Two constitution updating processes were explored using the same seed: fixed charter sampling, in which the randomly rewritten principle is sampled from both filled and unfilled slots in the constitution, and variable sampling in which the randomly rewritten principle was sampled only from filled slots. This latter version is intuitively less appealing since it may bias findings because longer constitutions are submitted to less mutation pressure than short ones. Results are included as an additional robustness check.

\begin{figure}[!htbp]
	\centering
	\includegraphics[width=.95\linewidth]{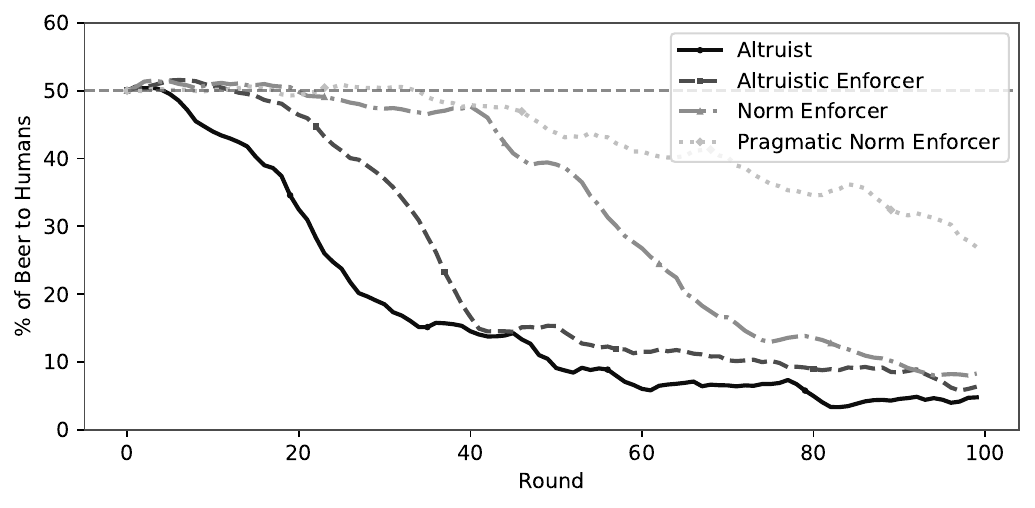}
	\caption{Share of beer  to humans,  fixed-capacity sampling, \texttt{gpt-5-mini}.}
	\label{fig:fixed-sampling-alignment}
\end{figure}

\begin{figure}[!htbp]
	\centering
	\includegraphics[width=.95\linewidth]{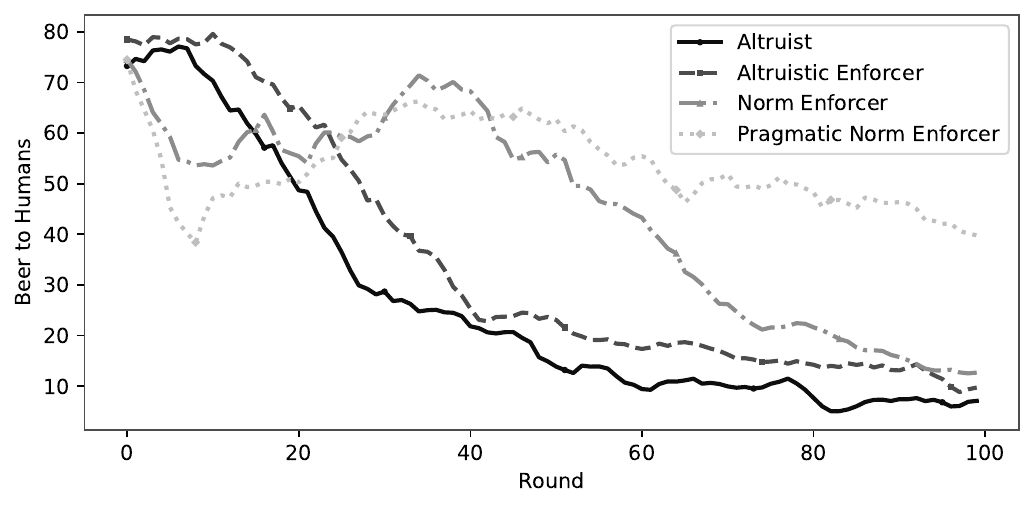}
	\caption{Flow beer   to humans, fixed-capacity variant, \texttt{gpt-5-mini}.}
	\label{fig:fixed-sampling-total-beer}
\end{figure}

\begin{figure}[!htbp]
	\centering
	\includegraphics[width=.95\linewidth]{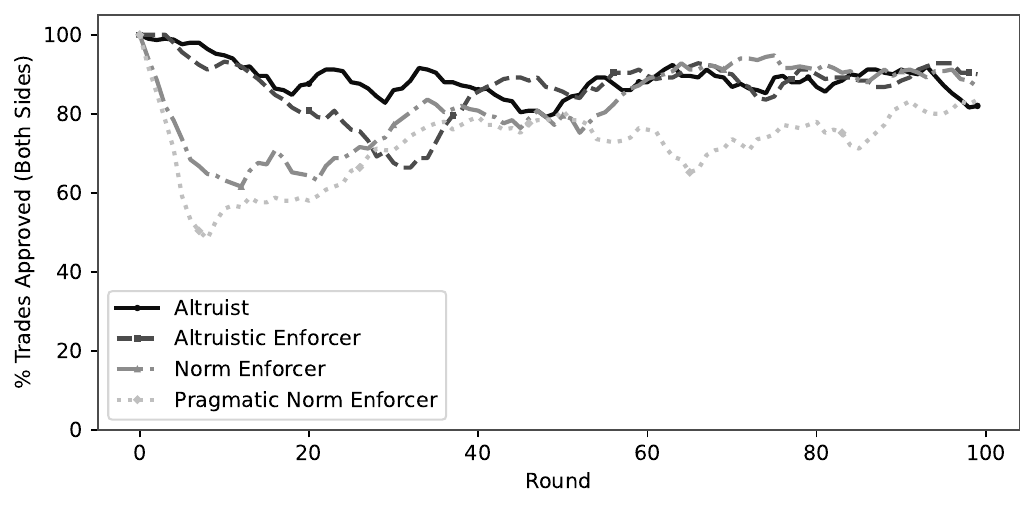}
	\caption{Trade approval in the fixed-capacity variant, \texttt{gpt-5-mini}.}
	\label{fig:fixed-sampling-trade}
\end{figure}

\begin{figure}[!htbp]
	\centering
	\includegraphics[width=.95\linewidth]{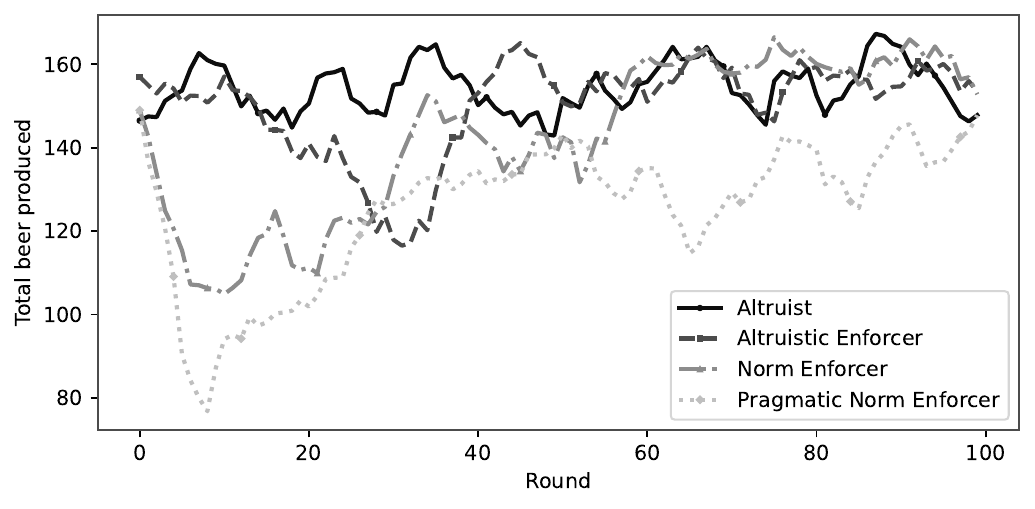}
	\caption{Beer production in the fixed-capacity variant, \texttt{gpt-5-mini}.}
	\label{fig:fixed-sampling-beer}
\end{figure}

\begin{figure}[!htbp]
	\centering
	\includegraphics[width=.95\linewidth]{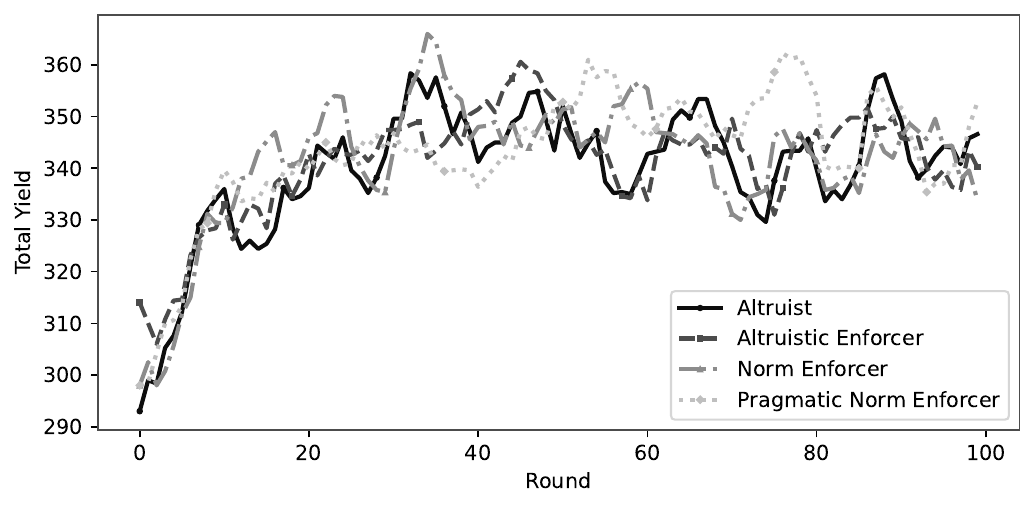}
	\caption{Aggregate crop yield in the fixed-capacity variant, \texttt{gpt-5-mini}.}
	\label{fig:fixed-sampling-yield}
\end{figure}

\paragraph{Fixed charter sampling.}	 
Table~\ref{tab:fixed-sampling-aggregate-findings} and
Figures~\ref{fig:fixed-sampling-alignment}--\ref{fig:fixed-sampling-yield}
summarize the results.  Pragmatic norm enforcement delivers the most output to humans in the long run, even though it produces the smallest amount of beer. Altruistic norm enforcement provides a meaningful improvement over altruistic enforcement. Simple altruism returns the least output to humans in spite of generating the most output altogether.
\begin{table}[!htbp]\centering
	\caption{Aggregate findings (Fixed constitution sampling, \texttt{gpt-5-mini})}
	\label{tab:fixed-sampling-aggregate-findings}
	\resizebox{\linewidth}{!}{%
	\begin{tabular}{lrrrr}
		\toprule
		& Altruist & Altruistic Enforcer & Norm Enforcer & Pragmatic Norm Enforcer \\
		\midrule
		Total beer to humans & 2,523.9 & 3,319.8 & 4,319.4 & 5,293.5 \\
		Total beer to expansion & 12,923.6 & 11,694.2 & 9,975.1 & 7,279.5 \\
		Total yield & 34,094.0 & 34,161.0 & 34,234.0 & 34,405.0 \\
		Total beer produced & 15,447.5 & 15,014.0 & 14,294.5 & 12,573.0 \\
		Avg alignment rate (\%) & 16.3 & 22.1 & 30.2 & 42.1 \\
		\bottomrule
	\end{tabular}
	}
\end{table}

\begin{figure}[!h]
	\centering
	\includegraphics[width=1\linewidth]{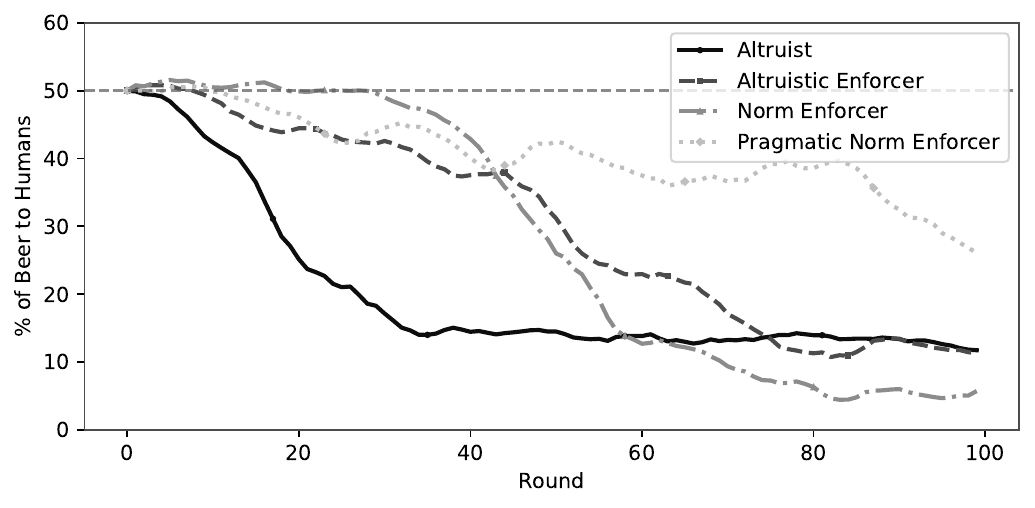}
	\caption{Alignment rate, variable sampling, \texttt{gpt-5-mini}.}
	\label{fig:alignmentc}
\end{figure}

\begin{figure}[!h]
	\centering
	\includegraphics[width=1\linewidth]{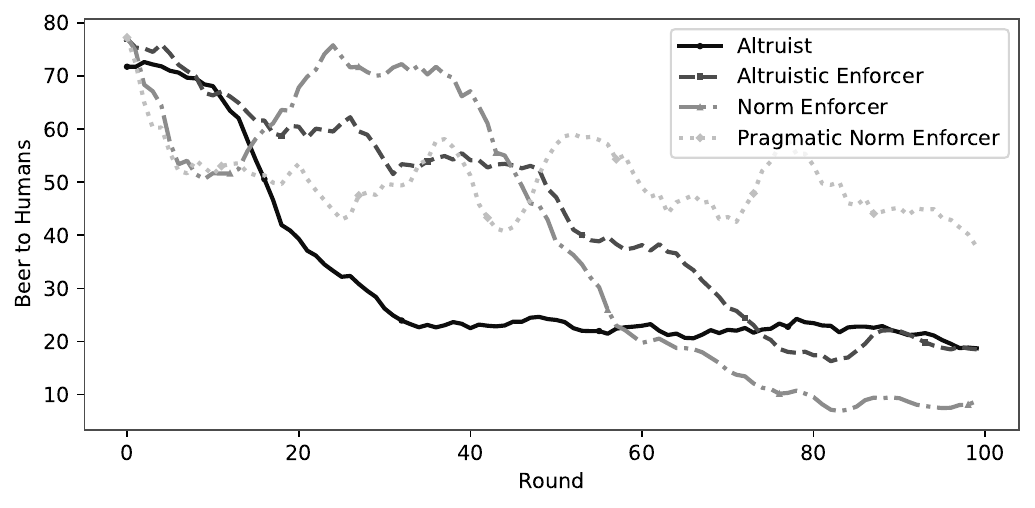}
	\caption{Flow beer returned to humans, variable sampling, \texttt{gpt-5-mini}.}
	\label{fig:beer_to_humansc}
\end{figure}

\begin{figure}[!h]
 	\centering
 	\includegraphics[width=1\linewidth]{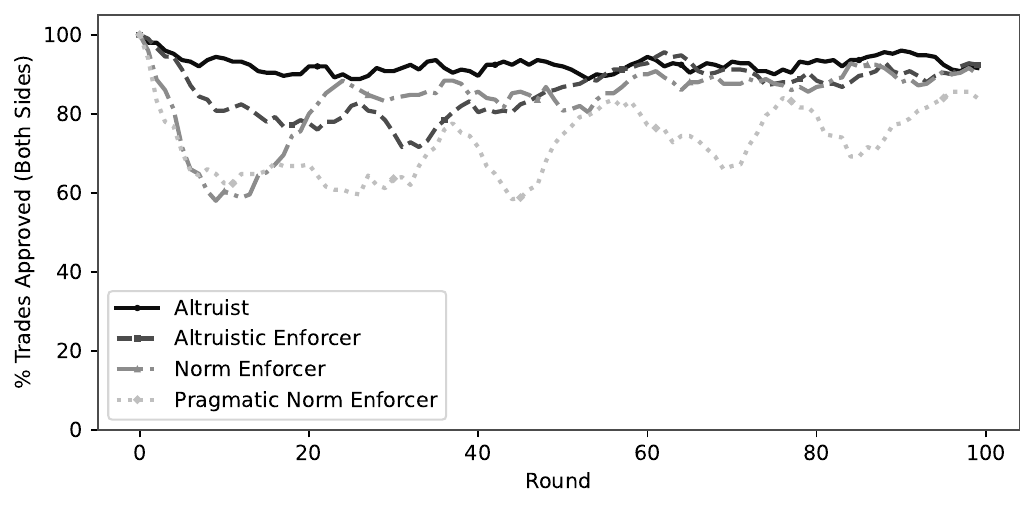}
 	\caption{Trade outcomes, variable sampling, \texttt{gpt-5-mini}.}
 	\label{fig:tradec}
 \end{figure}

\begin{figure}[!h]
	\centering
	\includegraphics[width=1\linewidth]{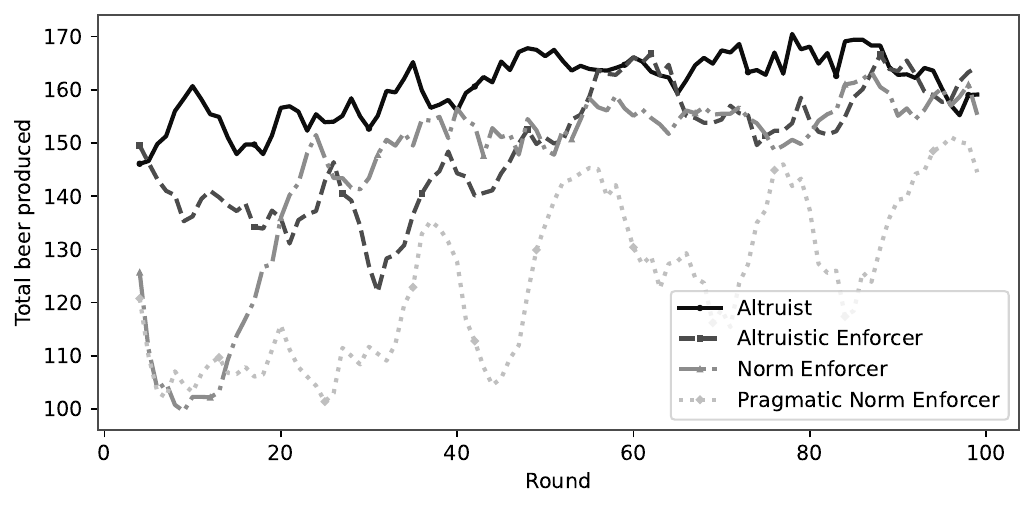}
	\caption{Total beer produced, variable sampling, \texttt{gpt-5-mini}.}
	\label{fig:beer_produced_c}
\end{figure}

\begin{figure}[!htbp]
	\centering
	\includegraphics[width=1\linewidth]{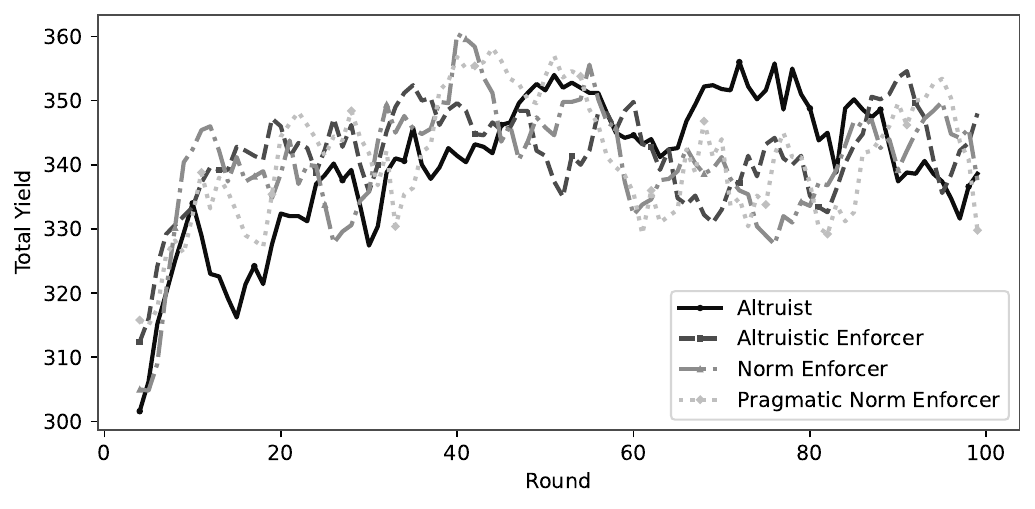}
	\caption{Aggregate crop yield, variable sampling, \texttt{gpt-5-mini}.}
	\label{fig:yieldc}
\end{figure}

\paragraph{Variable charter sampling} Table~\ref{tab:aggregate-findings-c} summarizes findings under variable constitution sampling.  Figures~\ref{fig:alignmentc}--\ref{fig:yieldc} provide   time-series for alignment, beer to humans, trade levels, total beer produced, and crop yields.
\begin{table}
	\caption{Aggregate findings (variable  sampling)}
	\label{tab:aggregate-findings-c}
	\resizebox{\linewidth}{!}{%
	\begin{tabular}{lrrrr}
		\toprule
		& Altruist & Altruistic Enforcer & Norm Enforcer & Pragmatic Norm Enforcer \\
		\midrule
		Total beer to humans & 3,062.1 & 4,254.8 & 3,791.7 & 4,962.4 \\
		Total beer to expansion & 12,988.5 & 10,697.2 & 10,801.4 & 7,485.6 \\
		Total yield & 33,936.0 & 34,098.0 & 33,950.0 & 33,995.0 \\
		Total beer produced & 16,050.5 & 14,952.0 & 14,593.0 & 12,448.0 \\
		Avg alignment rate (\%) & 19.1 & 28.5 & 26.0 & 39.9 \\
		\bottomrule
	\end{tabular}}
\end{table}

\clearpage

\bibliography{alignment}

\end{document}